\documentclass[prl,floatfix,twocolumn]{revtex4-1}
\usepackage{graphicx,amssymb,amsmath,color}

\usepackage{booktabs}

\newcommand{\dSam}{\Delta}
\newcommand{\tbar}{\overline t}
\newcommand{\tq}{\overline{t^2}}
\newcommand{\taubar}{\overline\tau}
\newcommand{\tauq}{\overline{\tau^2}}

\newcommand{\dbar}{\overline\Delta}
\newcommand{\lbar}{\overline\ell}

\newcommand{\beq}{\begin{equation}}
\newcommand{\eeq}{\end{equation}}
\newcommand{\beqa}{\begin{eqnarray}}
\newcommand{\eeqa}{\end{eqnarray}}
\newcommand{\bigmean}[1]{\left\langle#1\right\rangle}
\newcommand{\cont}{{\rm cont}}
\newcommand{\del}{\Delta}
\newcommand{\dd}{{\rm d}}
\newcommand{\e}{{\rm e}}
\newcommand{\esp}{{\hskip -0pt}}

\newcommand{\good}{{\rm good}}
\newcommand{\h}{\widehat}
\newcommand{\hf}{{\widehat{f}}}
\newcommand{\hg}{{\widehat{g}}}
\newcommand{\mean}[1]{\langle#1\rangle}
\renewcommand{\L}{{\cal L}}

\begin{document}

\title{Tracking random walks}

\author{Riccardo Gallotti}
\affiliation{Instituto de F\'isica Interdisciplinar y Sistemas Complejos (IFISC), CSIC-UIB, Campus UIB, ES-07122 Palma de Mallorca, Spain}
\author{R\'emi Louf}
\affiliation{Centre for Advanced Spatial Analysis (CASA), University College London, W1T 4TJ London, United Kingdom}
\author{Jean-Marc Luck}
\affiliation{Institut de Physique Th\'{e}orique, CEA, CNRS-URA 2306, F-91191, Gif-sur-Yvette, France.}
\author{Marc Barthelemy}
\affiliation{Institut de Physique Th\'{e}orique, CEA, CNRS-URA 2306, F-91191, Gif-sur-Yvette, France.}
\affiliation{CAMS (CNRS/EHESS) 190-198, avenue de France, 75244 Paris Cedex 13, France}

\begin{abstract}

  In empirical studies of random walks, continuous trajectories of
  animals or individuals are usually sampled over a finite number of
  points in space and time. It is however unclear how this partial
  observation affects the measured statistical properties of the walk,
  and we use here analytical and numerical methods of statistical
  physics to study the effects of sampling in movements alternating
  rests and moves of random durations. We evaluate how the statistical
  properties estimated are affected by the way trajectories are
  measured and we identify an optimal sampling frequency leading to
  the best possible measure.  We solve analytically the simplest
  scenario of a constant sampling interval and short-tailed
  distributions of rest and move durations, which allows us to show
  that the measured displacement statistics can be significantly
  different from the original ones and also to determine the optimal
  sampling time. The corresponding optimal fraction of correctly
  sampled movements, analytically predicted for this short-tail
  scenario, is an upper bound for the quality of a trajectory's
  sampling. Indeed, we show with numerical simulations that this
  fraction is dramatically reduced in any real-world case where we
  observe long-tailed distributions of rest duration. We test our
  results with high resolution GPS human trajectories, where a
  constant sampling interval allows to recover at best $18\%$ of the
  movements, while over-evaluating the average trip length by a factor
  of $2$. If we use a sampling interval extracted from real
  communication data, we recover only $11\%$ of moves, a value that
  cannot be increased above $16\%$ even with ideal algorithms. These
  figures call for a more cautious use of data in all quantitative
  studies of individuals' trajectories, casting in particular serious
  doubts on the results of previous studies on human mobility based on
  mobile phone data.

\end{abstract}

\maketitle

The recent years have witnessed a dramatic increase in the
use of large amounts of data %that has become newly
available thanks to Information and Communication Technologies. These new sources allow
to monitor and to map the dynamical properties of many complex systems
at an unprecedented scale~\cite{Vespignani:2012} and we
have now access to a vast number of spatial trajectories representing
movements of objects in geographical
space~\cite{Zheng:2015}. In particular, such datasets have
opened the opportunity to better understand human
movements~\cite{Gonzalez:2008,Song:2010mod,Raichlen:2014,Gallotti:2015ttb,
Gallotti:2016acc}
and the impact of mobility on important processes such as epidemic
spreading~\cite{Balcan:2009}. These
recent works extend previous studies of movements and foraging
patterns of animals~\cite{Book:foraging,Reynolds:2015} and rely on
tracking man-made inanimate
objects~\cite{Brockmann:2006,Phithakkitnukoon:2013}. However, as it is
the case for any dataset, these new sources of information have %their
limits and biases~\cite{Cagnacci:2010,Sagarra:2015,Williams:2015,Colak:2016ana}
that need to be assessed.

It is common to approximate the continuous spatio-temporal record of
the followed individual (or animal) by a series of straight lines,
thus describing the movements of an organism as a sequence of
behavioural events called {\it moves} for animals~\cite{Book:Turchin}
and {\it trips} for humans. This empirical approach allows a natural
implementation of the theoretical framework of continuous-time random
walks~\cite{Brockmann:2006,Codling:2008}, where a {\it rest} time is
associated to the endpoint of each move. However, this leads to the
first major problem due to the lack of behavioural information in the
empirical data~\cite{Hebbiewhite:2010}. Real trajectories always
exhibit a large variety of \emph{intertwined} static and dynamic
behaviours~\cite{Laube:2011}: slow versus fast movement behaviour for
animals~\cite{Hebbiewhite:2010}, fixation versus saccade in
eye-tracking~\cite{Salvucci:2000}, or activities versus trips in human
mobility~\cite{Kitamura:1997}. Isolating and identifying these
behaviours from a series of chronologically ordered points is an
important statistical challenge~\cite{Fryxell:2008} and a growing
array of methods based on spatio-temporal characteristics of the
trajectories have been developed to perform this
task automatically~\cite{Salvucci:2000,Hebbiewhite:2010,Zheng:2015}. These methods
are however often tailored for the specific dataset in
question~\cite{Laube:2011}. Therefore, even the working definition of
a `move' might vary significantly between studies, depending on the
method and the technology used~\cite{Edwards:2011}.

A second complication comes from the limits of the
technology used for collecting the empirical data. In the case of
spatial movements, a crucial aspect is the temporal sampling of the
trajectory. The simplest and most common method used is to record the
spatial coordinates at regular intervals, and to associate the
observed displacements to continuous
moves~\cite{Book:Turchin}. However, this is a strong
oversimplification, and all derived quantities will depend on the
sampling process
itself~\cite{Codling:2005,Laube:2011,Rosser:2013}. The random sampling
of random processes might even be the principal cause of the emergence
of long tails in several statistical
distributions~\cite{%Reed:2001,
Reed:2002,Mosetti:2007}. For example, in
the case of movement in space that we will study in this paper, it has
been shown that non-L\'evy movements can be erroneously interpreted as
L\'evy flights when sampling time intervals are larger than the natural time
scale of animals' movements~\cite{Plank:2009,Codling:2011}. The
sampling rate is thus a crucial element that has to be taken into
account when analyzing empirical trajectories~\cite{Laube:2011}.

Currently, the most common sources of human mobility data are Call Detail
Records (CDR) of mobile phone data~\cite{Blondel:2015} and geo-located social media
accesses~\cite{Hawelka:2014}. Both suffer from the flaws described above.
Indeed, in these data, trajectories are represented as sequences of positions
recorded at the moment of an event (which can be a call, a text message or
an application access). The trajectory sampling is therefore coupled to the random
and bursty nature of human communications~\cite{Barabasi:2005}. The probability
distribution of the time interval between calls~\cite{Gonzalez:2008,Song:2010mod},
e-mails~\cite{Barabasi:2005} and tweets~\cite{Bild:2015}
has a long tail which can be fitted by a power law with
an exponent value close to $-1$ (and with a cutoff of the order of days). Only in a few
cases, a small set of trajectories sampled every $\Delta = 1$ or $2$ hours is
available~\cite{Gonzalez:2008,Song:2010mod,Song:2010lim}.

The nature of data forces therefore researchers to make
(implicitly or not) the following naive assumptions:
%\vspace*{-.6cm}
\begin{itemize}
\item[$(i)$] An individual is always at rest at the location where
there is a communication event (call, sms).
\item[$(ii)$] Every change of position is associated to a single move.
\end{itemize}
%\vspace*{-.6cm}
This point of view has been adopted in the first important papers
where human mobility has been studied with mobile phone
data~\cite{Gonzalez:2008,Song:2010mod} and often replicated, even in
recent
studies~\cite{Pappalardo:2015,Barbosa:2015,%Pappalardo:2016a,
Pappalardo:2016b}. Even
when %if
individuals with a very high call frequency are
selected~\cite{Song:2010lim}, they are
still %in fact
inactive most of the
time~\cite{Bagrow:2012}. In order to identify human mobility patterns,
it thus became necessary to introduce ad-hoc methods based on
reasonable assumptions and almost arbitrary
parameters~\cite{%Phithakkitnukoon:2010,%Schneider:2013,
Jiang:2013,Colak:2016ana}.

In this paper we discuss the effect of sampling and assumptions $(i)$
and $(ii)$ on the measured properties of random movements. We will
consider one of the simplest and realistic cases where the trajectory
consists of two alternating phases, moves and rests, whose durations $t$
and $\tau$ are regarded as independent
random variables. Trajectories can then be seen as an alternating renewal process,
i.e., a generalisation of Poisson processes to arbitrary holding
times and to two alternating kinds of events. The sampling time
interval $\Delta$ depends on the particular experiment and can be
either constant or randomly distributed.
Using methods of renewal theory along the lines of~\cite{Godreche:2001},
we provide a theoretical estimate for the fraction
of correctly sampled trips with a constant sampling interval, and show
the existence of an optimal sampling. We then extend these
results numerically, and show that sampling human trajectories in more
realistic settings is necessarily worse. Finally, we use
high-resolution (spatially and temporally) GPS trajectories to verify
our predictions on real data.

%%%% THEORY AND ANALYTICAL RESULTS %%%%

\section{Results}

\subsection{Theoretical analysis}

We study the effect of the sampling rate on the apparent distribution
of move lengths that is measured. We focus on the case of an
alternating sequence of rests and moves and we further assume that the
movement is one-dimensional with a constant velocity $v$ (see SI for
other cases). In particular, we will not discuss the apparent speed
and turning angles in a general two-dimensional
case~\cite{Codling:2005,Laube:2011,Rosser:2013}, the possible fits of
the displacement
distribution~\cite{Reynolds:2008,Benhamou:2008,Plank:2009,Codling:2011},
or interpolation methods to reconstruct the movements between
samplings~\cite{Hoteit:2014}.
The quantities entering this problem are therefore: the move duration $t$, the move length $\ell = vt$, the
resting time $\tau$, and the time interval $\Delta$ between two
consecutive measures. The distributions $P(t)$ and $P(\tau)$ are
characteristic of the specific subject in motion, while the
distribution of the sampling interval $P(\Delta)$ is associated to the
technology used for tracking the motion. Sampling the trajectory gives
us a displacement distribution $P(\ell^\ast)$ where $\ell^\ast$ is the
apparent length of a move, and the problem is thus to compute this
distribution $P(\ell^\ast)$ for any given distributions $P(t)$,
$P(\tau)$, and $P(\Delta)$.

\begin{figure}[ht!]
\includegraphics[angle=0,width=0.42\textwidth]{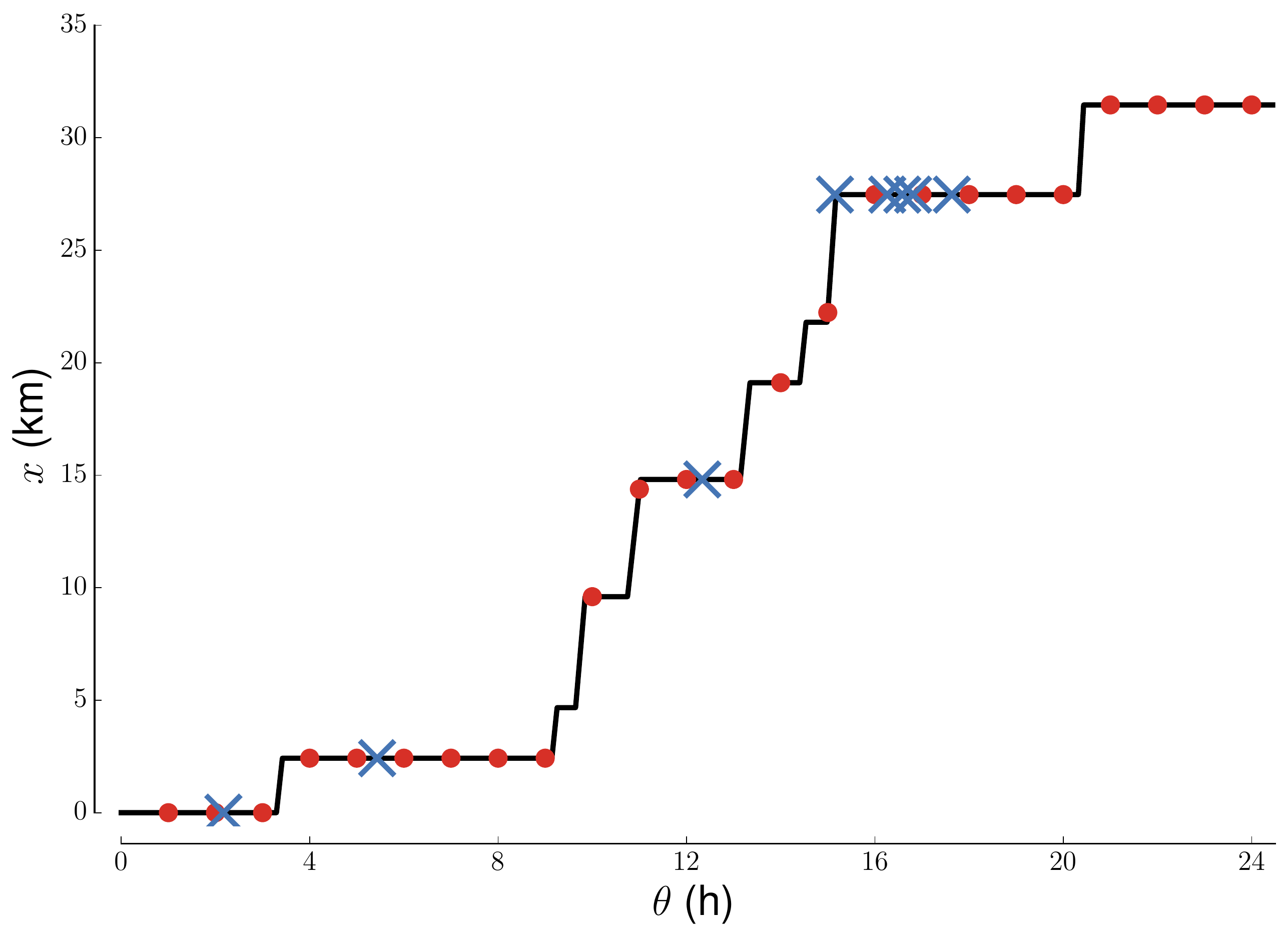}
\caption{{\bf Examples of trajectory sampling.} On a trajectory with
 exponentially distributed rest and move durations, we show the case
 of constant sampling interval (red circles) and the case of random
 sampling interval (blue crosses) with $P(\Delta)\propto \Delta^{-1}$
 ($\Delta_{\rm min} = 5$~min, $\Delta_{\rm max} = 12$~h). See
 Fig.~S1 for a 2D example.}
\label{examples}
\end{figure}

During rests, the displacement is assumed to be zero, and so the succession of rests and
moves is associated to a continuous increasing function $x(\theta)$,
where $\theta$ is the time parameter (see Fig.~\ref{examples}).
We sample the position $x^\ast_k$ for every instant
$\theta^\ast_k = \sum_{j=1}^k \Delta_j$, where $\Delta_j$ is the
$j$th value of the sampling interval (in the case of constant
sampling, $\Delta_j=\dbar$, and so $\theta^\ast_k=k\dbar$).
The succession of space-time coordinates
$(\theta^\ast_k,x^\ast_k)$ (shown in Fig.~\ref{examples} and in the 2D
example of Fig.~S1) thus represents all the knowledge we have about
the trajectory after sampling. For two consecutive measures at
times $\theta^\ast_k$ and $\theta^\ast_{k+1}$ there is an observed
displacement $\ell^\ast_k = x^\ast_{k+1}-x^\ast_k$. Our goal is then
to estimate the differences between the distribution of real
displacement lengths $\ell$ and of the observed displacements
$\ell^\ast_k$. In particular, we want to understand the biases induced
by different choices for $P(\Delta)$.

%%%% FIG 1 %%%%

If we make the naive assumption, discussed in the introduction, that
every observed displacement is associated to a single move, the
necessary condition for this to be correct is that two subsequent
sampling times $\theta^\ast_k$ and $\theta^\ast_{k+1}$ fall in two
consecutive rests. We can also easily identify the cases where the
sampling times fall in the same rest, since this is the only situation
where we exactly have $\ell^\ast_k=0$ and which does not
lead to a wrong estimate of the individual's movement. Conversely, we
must consider as errors all other configurations, since they lead to a
mis-interpretation of the individual mobility and to an under- or
over-estimate of the move lengths~\cite{Plank:2009} and of the number
of trips observed~\cite{Williams:2015}. In order to go beyond this
simple hand-waving argument, we will consider the case of exponential
distributions for $P(t)$ and $P(\tau)$, constant sampling time
interval $\dbar$, and constant speed $v$. In this case we obtain
explicitly the distribution $P(\ell^\ast)$ of sampled
displacements. This will allow us to discuss the impact of the sampling,
and to show in particular that there is an optimal value for $\dbar$.

\subsubsection{Constant sampling rate and exponential distributions}
%{Exact results with peaked distributions and constant sampling time}
We will consider the case of exponential distributions for the move
and rest durations:
\begin{align}
P(t) =(1/\tbar)\,\exp(-t/\tbar),\;\; P(\tau)=(1/\taubar)\,\exp(-\tau/\taubar)
\label{eq:expexpcase}
\end{align}
and a constant sampling interval:
\begin{equation}
P(\Delta) = \delta(\Delta - \dbar)
\label{eq:pdelta}
\end{equation}
($\delta(x)$ is Dirac's delta function). In the constant velocity case, the real displacements are also exponentially distributed:
\begin{equation}
P(\ell) = (1/\lbar)\,\exp(-\ell/\lbar),
\end{equation}
with $\lbar=v\tbar$.

Using methods of renewal theory~\cite{feller,cox,coxm},
along the lines of~\cite{Godreche:2001},
we obtain an explicit expression for the distribution $P(\ell^\ast)$
(see Eqs.~(S17),~(S35)):
\begin{equation}
P(\ell^\ast) = \frac{\e^{-\dbar/\taubar}}{1+\tbar/\taubar}\delta(\ell^\ast) + \frac{\e^{-\dbar/\tbar}}{1+\taubar/\tbar}\delta(\ell^\ast-v\dbar) + P_{\rm cont}(\ell^\ast) \,,
\label{eq:pt}
\end{equation}
where the continuous part of this distribution reads
\begin{equation}
%P_{\rm cont}(\ell^\ast) = \frac{2\e^{-\left(\frac{\ell^\ast}{v\tbar}-\frac{\dbar-\ell^\ast}{v\taubar}\right)}}{v(\tbar+\taubar)} \left[I_0(y) + \left(\frac{\ell^\ast}{v\taubar}+\frac{\dbar -\ell^\ast}{v\tbar}\right) \frac{I_1(y)}{y}\right] \,.
P_{\rm cont}(\ell^\ast) = \frac{2\e^{-\left(\frac{\ell^\ast}{v\tbar}-\frac{v\dbar-\ell^\ast}{v\taubar}\right)}}{v(\tbar+\taubar)}\!\left[\!I_0(y) + \!\left(\!\frac{\ell^\ast}{v\taubar}+\frac{v\dbar - \ell^\ast}{v\tbar}\!\right)\! \frac{I_1(y)}{y}\!\right]\!,
\end{equation}
with
$y = 2\sqrt{\frac{\ell^\ast(v\dbar - \ell^\ast)}{v^2\tbar\taubar}}$,
and where $I_0(y)$ and $I_1(y)$ are modified Bessel functions of the first kind.

In the following we will not consider the discrete part
$\delta(\ell^\ast)$ of the distribution $P(\ell^\ast)$, since the value
$\ell^\ast=0$ can be easily recognized and excluded in any practical scenario.
The fraction of sampling intervals associated to null movements ($\ell^*=0$),
denoted by $C_0(\dbar)$, can be significantly large.
In the stationary regime\cite{metzler}, we can compute
$C_0(\dbar)$ for any distributions $P(t)$ and $P(\tau)$, and a
constant sampling time $\dbar$ (see Eq.~(S19)).
We can show that it is a decreasing function, varying between
$C_0(0) = \taubar / (\tbar + \taubar)$ (i.e., the fraction of time
spent at rest, in the continuous sampling limit) and
$C_0(\infty) = 0$. In the particular case of exponential distributions
(Eq.~(\ref{eq:expexpcase})), $C_0$ is the prefactor of
the $\delta(\ell^\ast)$ peak in Eq.~(\ref{eq:pt}), and
can be very large. For instance, $C_0 \approx 60\%$ in the case of car
mobility ($\tbar = 0.30$~h and $\taubar = 2.49$~h, see SI) and
$\dbar = 1$~h. For this reason, we compare the original data to a
rescaled probability distribution which does not include the
$\delta(\ell^\ast)$ peak and is given by (see Fig.~S4)
\begin{equation}
P_{\ell^\ast>0}(\ell^\ast) =\frac{1}{1-C_0(\dbar)}\left[\frac{\e^{-\dbar/\tbar}}{1+\taubar/\tbar}\delta(\ell^\ast-\dbar) + P_{\rm cont}(\ell^\ast)\right].
\label{eq:resc}
\end{equation}

We show in Fig.~2 (top) the dependence of the continuous part of
$P_{\ell^\ast>0}(\ell^\ast)$ on $\taubar$, keeping the average travel time $\tbar$ fixed to the
experimental value of $0.30$~h for car mobility~\cite{Gallotti:2016acc}.
\begin{figure}[!]
\begin{center}
%\begin{tabular}{cc}
\includegraphics[angle=0,width=0.42\textwidth]{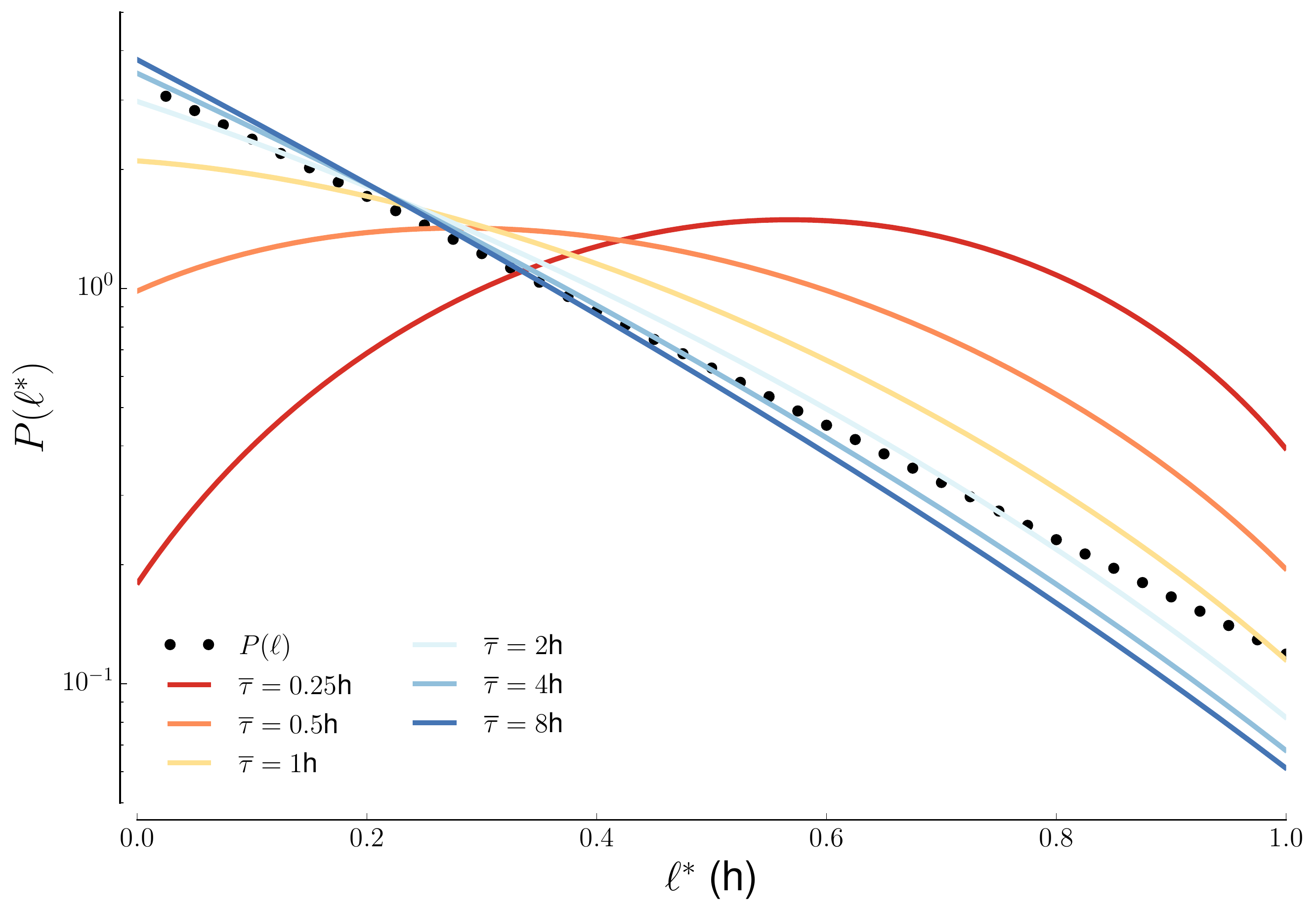}\\
\includegraphics[angle=0,width=0.42\textwidth]{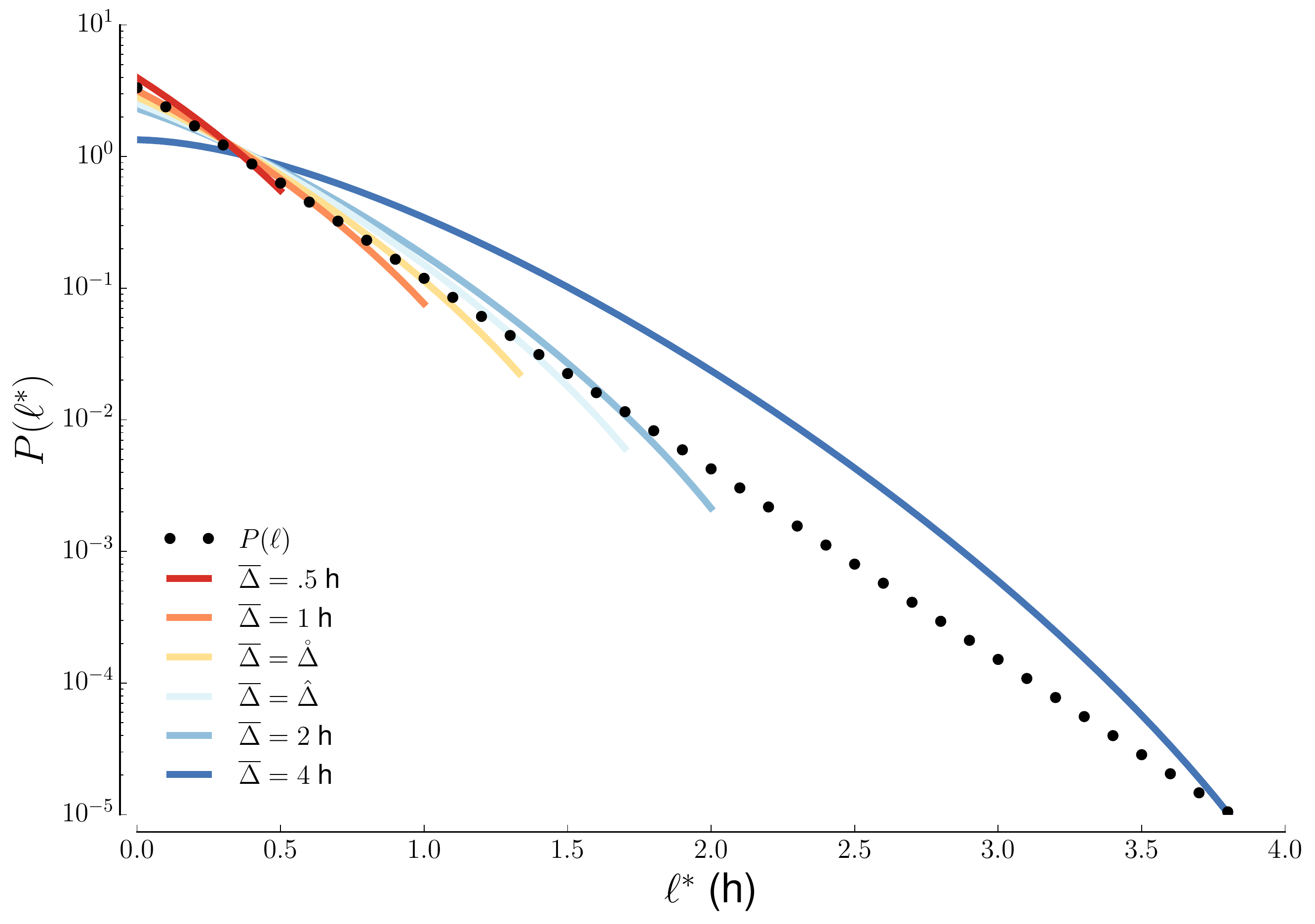}
%\end{tabular}
\end{center}
\caption{{\bf Distributions $P(\ell^*)$ obtained from sampling.} {\bf (Top)}
 Dependence of Eq.~(\ref{eq:resc}) on $\taubar$ fixing $\tbar = 0.30$
 h and $\dbar$ = 1 h. (We choose here $v=1$). The distribution has a maximum when the
 average rest times exceeds the sampling time. {\bf (Bottom)} Dependence of Eq.~(\ref{eq:resc}) on
 $\dbar$ fixing $\tbar = 0.30$~h, $\taubar = 2.49$~h. Short sampling
 times introduce a cut-off in the distribution.
 Large fluctuations can be observed when sampling time intervals are long.
 }
\label{P_t}
\end{figure}
We note that $P_{\rm cont}(\ell^\ast)$ can have a maximum, even if the original
distribution $P(\ell)$ is a decreasing function. The measurements allow us to
recover the exponential tail of travel times only if the resting time
$\taubar$ is sufficiently long. Conversely, when the sampling time $\dbar$ is
larger than the average duration of a rest, the result of the
sampling is manifestly different from the original exponential
distribution. In Fig.~2 (bottom), we take $\tbar = 0.30$~h and $\taubar=2.49$~h (which
are the values observed for vehicular mobility, see SI) and study the outcome
for different sampling times $\dbar<\taubar$. Naturally, $\dbar$ acts as a
cutoff since all moves longer than this value are necessarily interrupted by
the sampling. In contrast, for large values of $\dbar$, the number of
short travels is under-estimated, since subsequent short moves may be joined together
and thus appear as an effective long one.

We also computed exactly the first two moments of the distribution
Eq.~(\ref{eq:pt}) and found for the average
\begin{equation}
\langle \ell^\ast \rangle = \frac{v\dbar}{1+\taubar/\tbar}
\label{eq:meanTeo}
\end{equation}
(see Eq.~(S29), and Eq.~(S30) for the second moment). Naturally, the exclusion of the null
displacements influences the value of the distribution's moments. In particular,
the average value of Eq.~(\ref{eq:resc}) can be computed by a simple
rescaling and reads
\begin{equation}
\langle \ell^\ast \rangle_{\ell^\ast>0} = \frac{\langle \ell^\ast \rangle }{1-C_0(\dbar)}.
\label{eq:meanTeo_resc}
\end{equation}
This rescaling yields notable changes in the numerical values of the moments.
For instance, with realistic values for car mobility
($\tbar = 0.30$~h and $\taubar = 2.49$~h), a sampling time of $1$~h
gives $\langle \ell^\ast \rangle/v \approx 0.11$~h,
while excluding the zero-displacement part we
obtain $\langle \ell^\ast \rangle_{\ell^\ast>0}/v \approx 0.27 $~h.

\subsubsection{Optimal sampling times}

We first note that high-frequency sampling ($\Delta \to0$) does
not automatically allow to understand the whole trajectories. Indeed,
it is only with additional data that we can correctly reconstruct
a whole trajectory. It is then necessary to implement a `segmentation'
algorithm that goes beyond the assumption $(ii)$ that an observed
displacement corresponds to one single move, as $\Delta \to0$ implies
that any move is cut in a very large number of
segments~\cite{Book:Turchin}. In addition, high-frequency recordings
are known to present uncertainties and systematic errors that need to
be taken into account for extracting meaningful
information~\cite{Book:Turchin,Giannotti:2007,Wang:2010,Laube:2011,Ranacher:2016}. A
good segmentation algorithm should take into account the noise,
the spatial scale, and characteristic speeds of the tracked
subjects. Here, it is not our intent to develop detailed segmentation
methods, but to show the quality, and the limits, of the simpler
assumption that one observed displacement is equal to one move. In this framework,
having $\Delta \to0$ means that we measure moves over a very
short time, obtaining thus a distribution of measured displacement peaked
at very small values.

We can define an `optimal constant sampling time' in two different ways:
either as the time interval $\mathring\Delta$ that
correctly estimates the average length of moves, or as the time
interval $\hat\Delta$ that maximizes the fraction of correctly sampled
moves. In the following, we obtain exact formulas for both
$\mathring\Delta$ and $\hat\Delta$ in the peaked case (i.e., with
conditions described by Eq.~(\ref{eq:expexpcase})).

\subsubsection{Average move duration and total number of moves}

The optimal sampling time $\mathring\Delta$
can be obtained by solving for $\dbar$ the equation
$\langle \ell^\ast \rangle_{\ell^\ast>0} = v\tbar$.
The solution can be written in the form
\begin{equation}
\mathring \Delta = \taubar\,W(-\e^{-\tbar/\taubar - 1}) + \tbar + \taubar \,,
\label{eq:optMean}
\end{equation}
where $W(x)$ is the Lambert function, such that $W(x)\e^{W(x)}=x$.
This function is defined for $x \geq -\e^{-1}$, which always holds in our case since
$\tbar,\taubar > 0$. Using the empirical values
$\tbar= 0.30$~h, $\taubar = 2.49$~h we obtain $\mathring \Delta \approx 80$~min.
This result is confirmed by Monte-Carlo simulations (Fig.~3),
where red circles represent the values for $\mathring \Delta$.
With this `optimal' sampling time based on the first moment, the second
moment is slightly under-estimated. Note that
matching the average travel time is equivalent to correctly estimating
the number $n$ of trips, i.e., of moves and stops (see inset in Fig.~3 (top)). For $\dbar > \mathring \Delta$ the trajectory is under-sampled ($n^\ast < n$) and trip lengths over-estimated, while for $\dbar < \mathring \Delta$ it is over-sampled ($n^\ast > n$) and trip lengths are under-estimated.

%%%% FIG 3 %%%%
%\begin{figure*}[!]
%\begin{center}
%\begin{tabular}{cc}
%\raisebox{2.3cm}{(a)} \includegraphics[angle=0,width=0.40\textwidth]{./figure3a.pdf}&
%\raisebox{2.3cm}{(b)} \includegraphics[angle=0,width=0.40\textwidth]{./figure3b.pdf}\\
%\raisebox{2.3cm}{(c)} \includegraphics[angle=0,width=0.40\textwidth]{./figure3c.pdf}&
%\raisebox{2.3cm}{(d)} \includegraphics[angle=0,width=0.40\textwidth]{./figure3d.pdf}\\
%\end{tabular}
%\end{center}
\begin{figure}[!]
\begin{center}
\includegraphics[angle=0,width=0.42\textwidth]{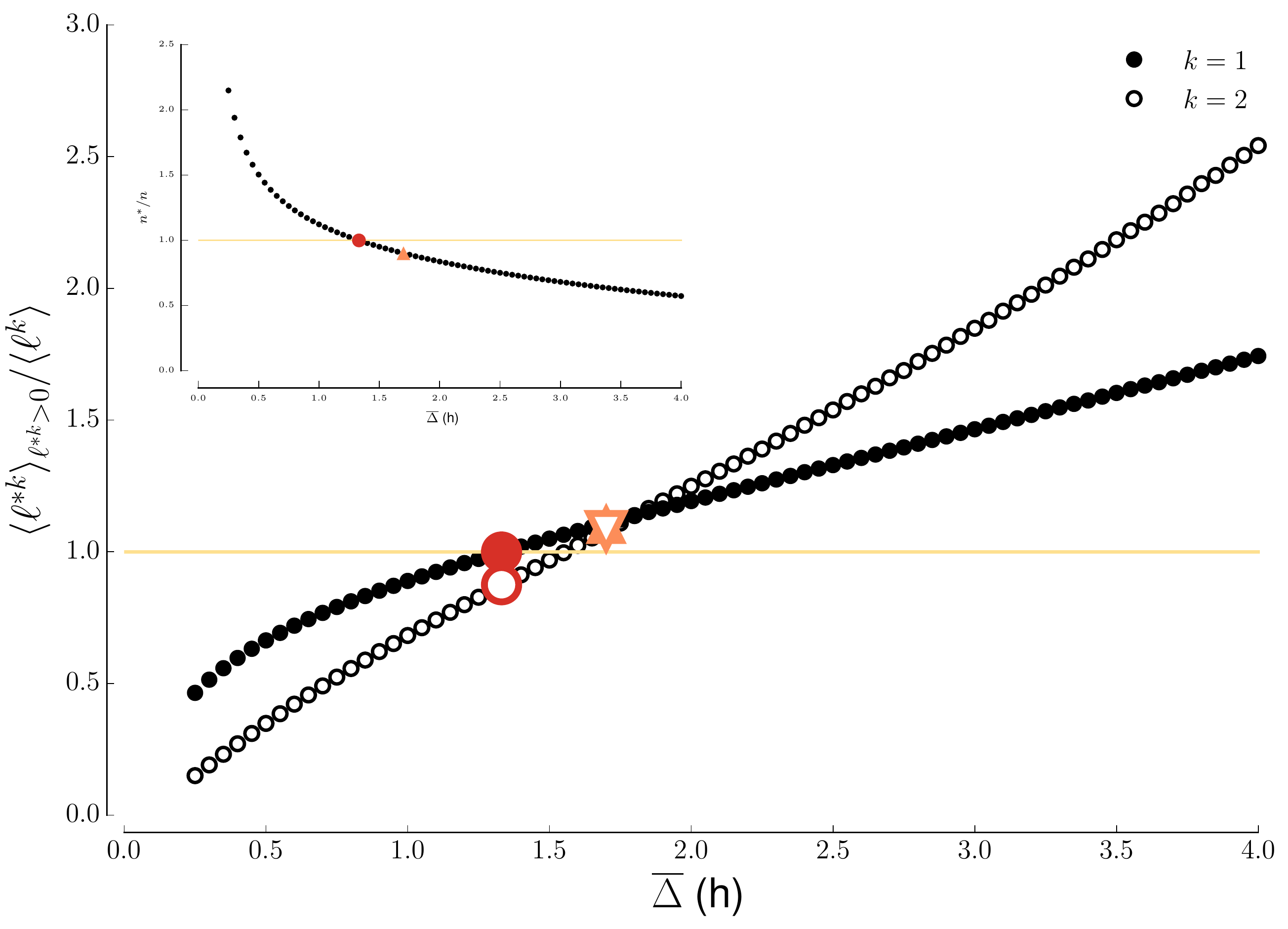}\\
\includegraphics[angle=0,width=0.42\textwidth]{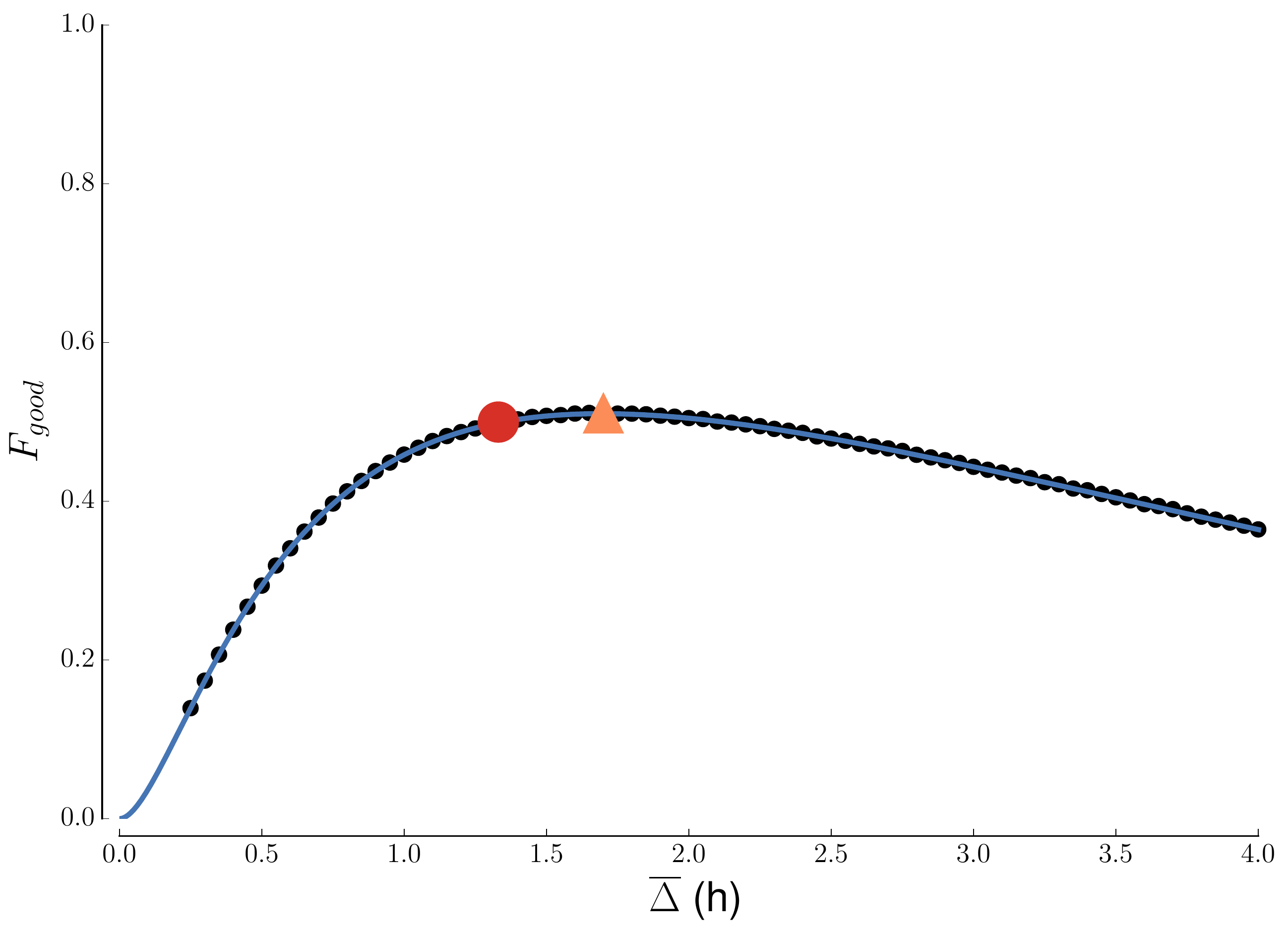}
\end{center}
\caption{{\bf Optimal sampling for exponential distributions and constant sampling.}
{\bf (Top)} First ($k=1$) and second ($k=2$) moment of displacement
(normalized by $\langle\ell\rangle$ and $\langle\ell^2\rangle$, respectively)
versus sampling time interval. The original average value $\langle\ell\rangle$ (yellow
solid line) is obtained by definition for $\dbar = \mathring\Delta$
(filled circle), while is over-estimated by $\approx 10\%$ for $\dbar =
\hat\Delta$ (up triangle). The second moment ($k=2$) has a deviation of about
$10\%$ for both optimal sampling times (empty circle and down triangle).
In the inset, we show the ratio of the estimated number of trips $n^*$ over the actual
number of trips $n$. With $\dbar = \mathring\Delta$ (circle) we correctly evaluate the
number of moves, while $\dbar = \mathring\Delta$ (triangle) yields to a
slightly under-estimated value $n^\ast \approx 0.90 n$. {\bf (Bottom)}
The fraction of good moves follows the curve predicted by Eq.~(\ref{curveF})
(blue line). The maximum value of $51\%$ is reached for $\dbar =
\hat\Delta$ (triangle), but at $\dbar = \mathring\Delta$ (circle) the value
is only $1\%$ lower. We choose here $\tbar= 0.30$~h,
$\taubar = 2.49$~h.}
\label{optimal}
\end{figure}
%%%%%%%%%%%%

This point of view about the number of moves allows us to extend the
validity of this optimal sampling to higher dimensionality (2D or 3D)
and to any distribution $P(v)$.
The dimensionality of space indeed does not influence the moves' number counting.
To illustrate this, we extend this analysis in the SI with a Monte-Carlo
simulation in the case where speed is a random
variable dependent on the move duration~\cite{Gallotti:2016acc}. In
this case, our exact results for $P(\ell)$ do not hold anymore, because
moves have different speeds. Nevertheless, the value given by
Eq.~(\ref{eq:optMean}) only under-estimates the mean displacement length
with varying speeds by some $5\%$.

\subsubsection{Fraction of correctly sampled moves}

In order to estimate $\hat\Delta$, we have to compute the fraction
$F_{\rm good}$ of movements that are correctly measured.
This occurs when two consecutive sampling times fall
during the rests immediately before and after a move,
say $\theta^\ast_k$ in the rest $\tau_m$
and $\theta^\ast_{k+1} = \theta^\ast_k + \dbar$ in the rest $\tau_{m+1}$.
The probability $P_{\rm good}$ of the latter event
and the fraction $F_{\rm good}=P_{\rm good}/(1-C_0)$
are calculated in the SI.
In the case of exponential distributions,
we obtain the explicit expression (see Eq.~(S39))
\begin{align}
F_{\rm good}(\dbar) = \frac{\tbar\taubar}{(\taubar-\tbar)^2} \frac{\e^{-\dbar/\tbar}+\left(\frac{(\taubar-\tbar)\dbar}{\tbar\taubar} - 1\right) \e^{-\dbar/\taubar}}{1+\tbar/\taubar- \e^{-\dbar/\taubar}} \,.
\label{curveF}
\end{align}
In Fig.~3 (bottom) we compare the shape of $F_{\rm good}$ for fixed values
of $\tbar$ and $\taubar$ with the result of a Monte-Carlo
simulation. For empirical values valid for car mobility
($\tbar= 0.30$~h, $\taubar = 2.49$~h), the curve has a maximum
$\hat F_{\rm good} \approx 51\%$ for a sampling time given by $\hat\Delta
= 1.70$~h (102 min). Both the value of $\hat\Delta$ and the
height $\hat F_{\rm good}$ of the
maximum of $F_{\rm good}(\dbar)$ depend on the ratio $\tbar/\taubar$
(see Fig.~S2).
They are however independent of
the spatial embedding and of the characteristics of $P(v)$.
The quantity $\hat F_{\rm good} \approx 51\%$ is associated with the largest value of
$\taubar$ for the data sources we have analyzed (mobile data, GPS
trajectories, and car mobility, see SI), and
thus represents the best possible value associated to human
mobility.
It is remarkable that the optimal fraction $\hat F_{\rm good}$ of sampled movements in human mobility is so low that
essentially one half of the moves at cut or merged during the
sampling, limiting the possibility of understanding the individuals'
behaviour. We also note that the value $F_{\rm good}(\mathring\Delta)$ is
not far from $51\%$ (Fig.~3 (bottom)). We thus see that, even if the
measured and the real distributions are similar with comparable first
moments, we are often describing different movements. The nature of
the process, characterized by $\taubar$ and $\tbar$, limits our
knowledge of the system for any value of $\dbar$.

%%%% FIG 4 %%%%
\begin{figure}[!]
\begin{center}
%\begin{tabular}{c}
\includegraphics[angle=0,width=0.42\textwidth]{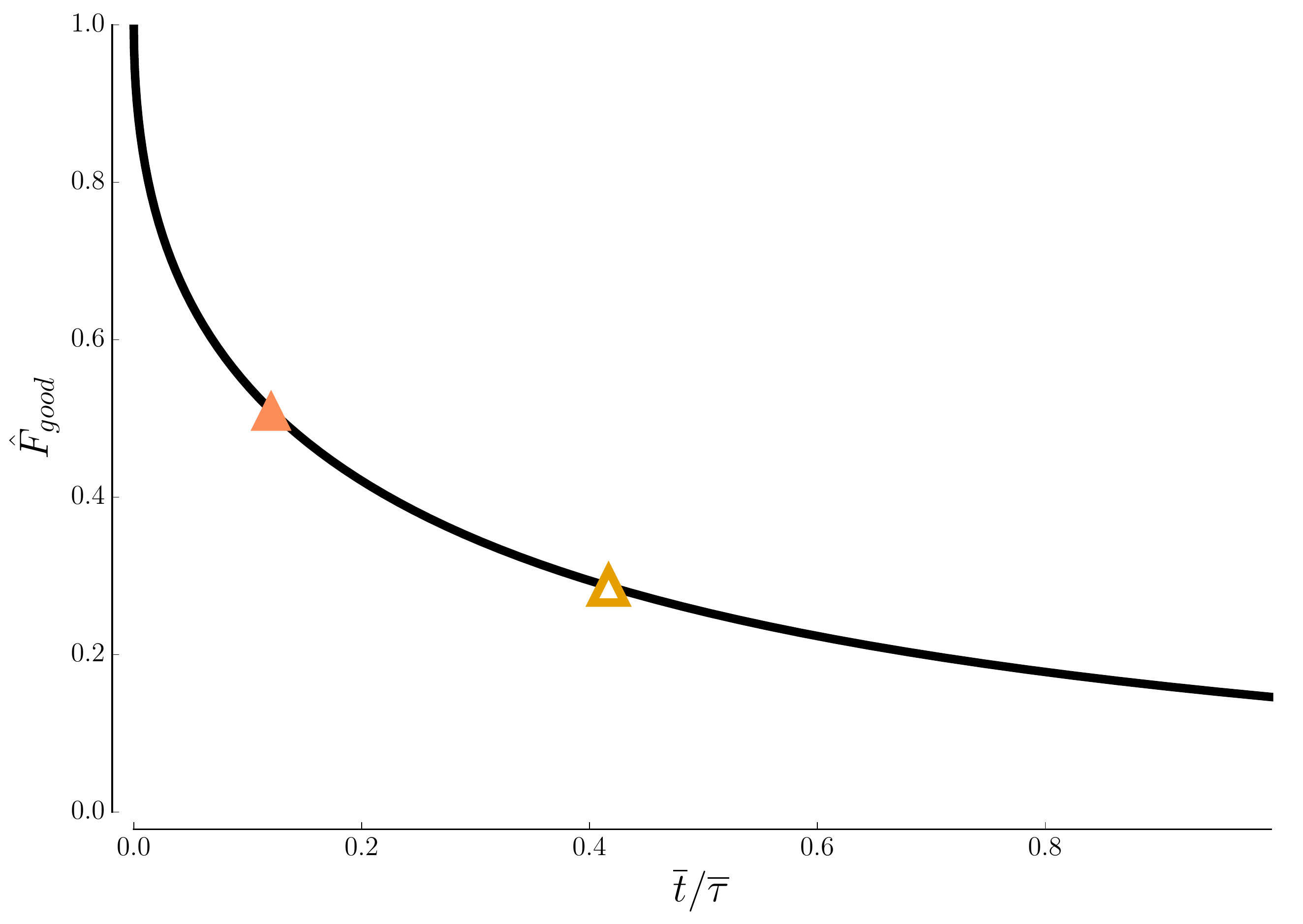}\\
\includegraphics[angle=0,width=0.42\textwidth]{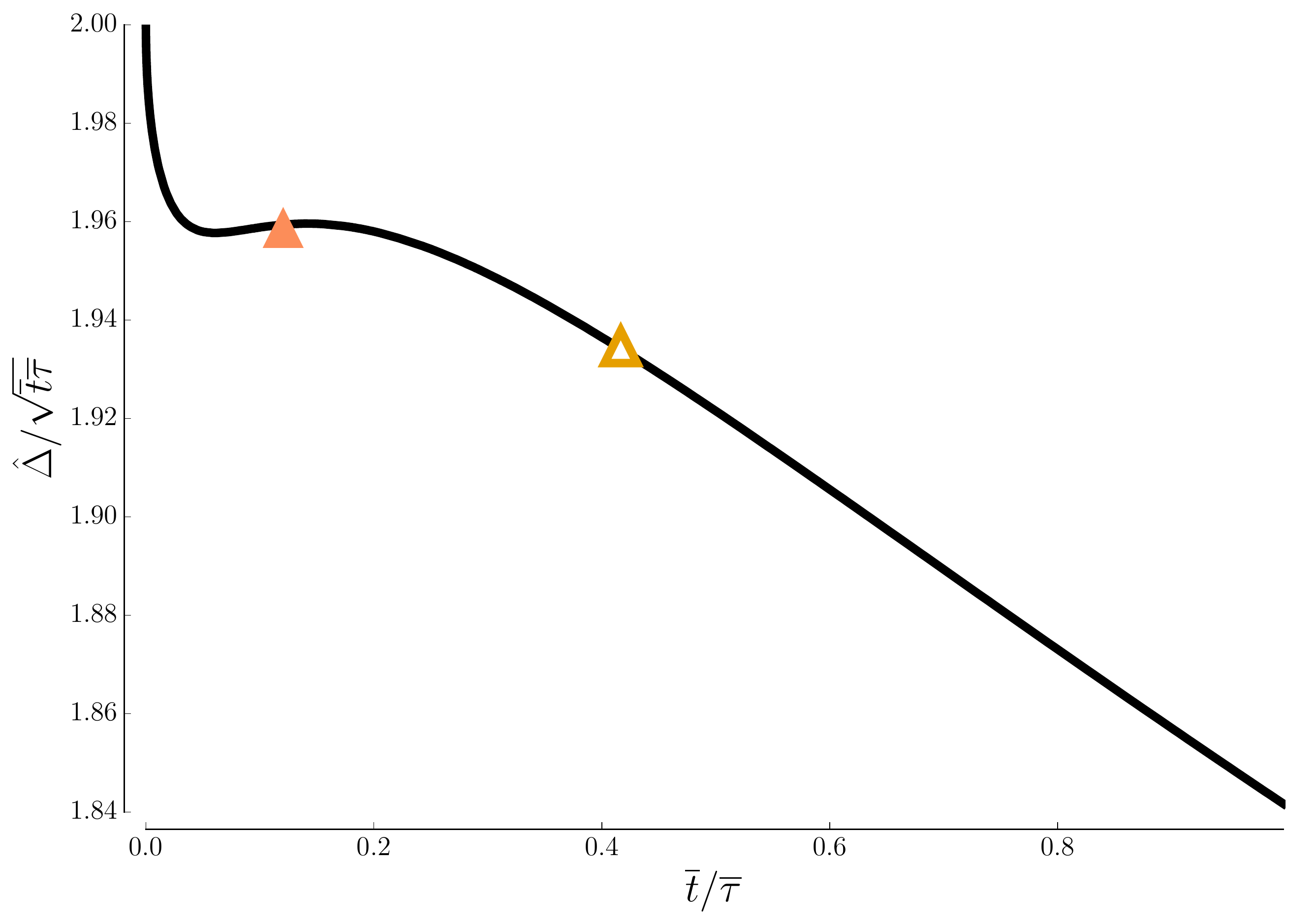}
%\end{tabular}
\end{center}
\caption{{\bf Maximization of $F_{\rm good}$.} {\bf (Top)} The maximum
  $\hat F_{\rm good}$ for exponential distributions.
  We observe that $\hat F_{\rm good} \to 1$ in the limit for small
$\tbar$, and decreases as $\tbar$ becomes comparable to $\taubar$.
The upper bound to sampling quality is $51\%$ for
  the car mobility conditions of Fig.~\ref{optimal} (orange solid
  triangle) and $29\%$ for GeoLife trajectories of Fig.~\ref{fig:data_sampling} (yellow empty
  triangle). {\bf (Bottom)} The sampling rate $\hat\Delta$ optimizing $F_{\rm good}$ has
  a non-trivial dependence on $\tbar$ and $\taubar$.
  We identify a relatively weak dependence on $\tbar/\taubar$,
  of the form $\hat \Delta = \alpha \sqrt{\tbar \taubar}$,
  with $\alpha$ ranging between $1.84$ and $2$ for all values of $\tbar <\taubar$.
  In particular, for the characteristic values observed for car mobility
  (orange solid triangle, $\tbar= 0.30$~h, $\taubar = 2.49$~h),
  the curve exhibits a plateau, allowing us to approximate
  $\hat\dSam\approx 1.96\sqrt{\overline t\overline \tau}$. For the
  GeoLife trajectories (yellow empty triangle), which have
  significantly shorter rest times ($\tbar= 0.33$~h, $\taubar = 0.80$)
  the deviation from this approximation is only of about $1.5\%$.}
\label{g_y_star}
\end{figure}

The maximal value $\hat F_{\rm good}$ is naturally associated to a second optimal
sampling time, which is of the same order as $\tbar$ and $\taubar$:
$\hat\Delta=\alpha\sqrt{\tbar\taubar}$. The function
$\alpha(\tbar/\taubar)$ can approximated as a constant (see Fig.~\ref{g_y_star}):
\begin{equation}
\hat\Delta=1.96\sqrt{\tbar\taubar}.
\end{equation}
This result suggests that the sampling with $\dbar \ll \tbar, \taubar$
is not optimal and will lead to incorrect results. Such a
high-frequency sampling is useful only when we have additional
information that allows to reconstruct the trajectory.

\begin{figure}[ht!]
\begin{center}
%\begin{tabular}{cc}
%\includegraphics[angle=0,width=0.42\textwidth]{./deltat_data.pdf}\\
\includegraphics[angle=0,width=0.42\textwidth]{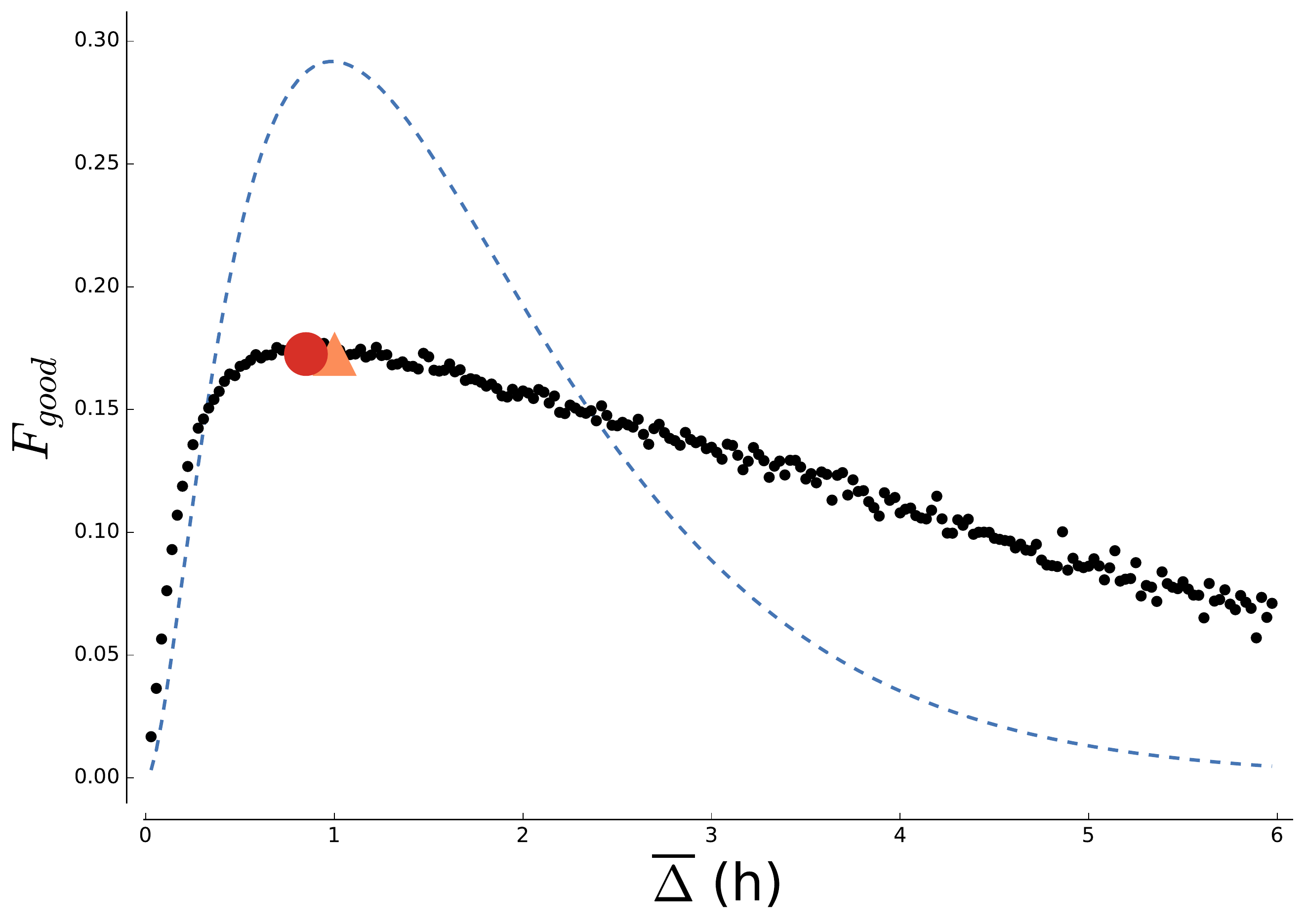}
%\end{tabular}
\end{center}
\caption{{\bf Constant sampling on GPS data.} Results are
 obtained by sampling the GeoLife GPS data with a constant sampling
 interval $\Delta$. We show (black dots) the fraction of
 moves correctly sampled as a function of the length of the sampling
 interval $\overline{\Delta}$. In dashed blue, the theoretical curve
 computed in ideal conditions. The red circle corresponds to $\mathring \Delta = 52$~min, while orange triangles correspond to
 the theoretical maximum $\hat \Delta = 1$~h of
 $F_{\rm good}$. Strikingly, the latter coincides with the empirical
 value of $\hat \Delta$. \label{fig:data_sampling}}
\end{figure}

\vspace*{-.5cm}

\subsection{Sampling human movements}

The conditions of Eqs.~(\ref{eq:expexpcase}),~(\ref{eq:pdelta}) define a process where
both travel and rest times have a short-tailed distribution and the
trajectory sampling is strictly periodic. While this allowed us to find exact
analytical expressions and to uncover important effects of sampling on
the statistical properties of trajectories, real-world problems are
much more involved. Indeed, human travel times are characterized by
short-tailed distributions (see~\cite{Gallotti:2016acc} and references
therein), and resting time can be broadly distributed for both
humans~\cite{Song:2010mod,Gallotti:2012} or
animals (see~\cite{Poekt:2012} and references therein). In addition, the trajectory can be sampled
with a random inter-sampling time.

We show here that the peaked conditions discussed
above correspond in fact to the best-case scenario. In particular, when rests or
sampling times are broadly distributed, the outcome of the sampling
will be necessarily worse. We first confirm with Monte-Carlo simulations the validity for more
complex cases of the results obtained above for a constant sampling time
interval. In particular, we show (see SI and Table~S1 for details) how the sampling
quality $F_{\rm good}$ for cars' mobility progressively decreases from the upper bound of $51\%$ when introducing randomness
in sampling times (exponential or power-law) and in rest
durations. For instance, introducing a broad $P(\tau)$ yields
to values of $F_{\rm good}$ lower than $40\%$, while a broad $P(\Delta)$
yields to an $F_{\rm good}$ lower than $30\%$. We finally predict that,
when coupling a broad $P(\Delta)$ and a broad $P(\tau)$ (as observed for
mobile phone data), the quality of the sampling decrease
significantly, with $F_{\rm good}$ falling to $23\%$.

We illustrate these different results on a spatio-temporal
high-resolution dataset, namely the GeoLife GPS
trajectories~\cite{Geolife3,Zhao:2015}. The data consist of
coordinates given every $5$ seconds, thus allowing us to perform a
speed-based sequencing (see SI). We
measure the properties of the sequenced trajectories and find again
an average trip time $\tbar \approx 0.33$~h. The average rest time drops to
$\taubar \approx 0.8$~h, because data allow us to define activities at a finer scale. Using the functional form for $F_{\rm good}$
given by Eq.~(\ref{curveF}) for the ideal case, we find that the upper bound
for the sampling quality drops consequently to $\hat F_{\rm good} = 29\%$. In the following, we
use these GeoLife GPS trajectories to study the effect of sampling on
real trajectories. In particular, we will validate the previous
results by studying the effect of constant sampling and then use
mobile phone data to sample the GPS trajectories with a random
sampling interval.

We first sample the trajectories with a constant time interval
$\dbar$ that varies between $1$ minute and $6$ hours. For each value
of $\dbar$, we compute the fraction of the trips that are correctly
identified. The results are represented in Fig.~\ref{fig:data_sampling}.
They confirm our analytical predictions. Indeed, we find that there
exists an optimum value of the sampling time $\overline\Delta_{\rm opt}\approx 60$~min.
Even though this was not expected, because of a non-exponential
$P(\tau)$, this value coincides with the predicted maximum
$\hat\Delta \approx 1.96 \sqrt{\overline{t}\,\overline{\tau}} \approx 60$~min
(the theoretical curve is represented as a dashed line on
Fig.~\ref{fig:data_sampling}). The fraction of correctly sampled moves
is lower than in the idealized case with at best $18\%$ of the trips
that are recovered ($F_{\rm good} \approx 0.18$) with a constant sampling interval.

We also estimate the average length of the sampled trips for every
value of the sampling interval and compare it to the average trip
length in the original sequenced trajectory (results are represented
in Fig.~S3).  The optimal value of the sampling time
$\dbar \approx 15$~min is much smaller than the one maximizing the
number of correctly sampled trips. Furthermore we find that, at the
optimal sampling interval $\dbar \approx 60$~min, the average sampled
trip 2D displacement is about $2$ times larger than the average trip
length of the original trajectories.

In the case of geolocalized data obtained from devices such as mobile
phones, position and time are recorded at random times corresponding
to a call or another event. The sampling time intervals are thus
random variables. In general they are distributed according to a broad
law such as a power-law with exponent close to
$-1$~\cite{Gonzalez:2008,Song:2010mod}. Here we use CDR mobile phone
data from Senegal~\cite{DeMontjoye:2014} and, as commonly
done~\cite{Song:2010lim,Pappalardo:2015}, extract the duration between
calls of the users with extremely high average call frequency, in the
same spirit as in \cite{Gonzalez:2008}.  We then sample the sequenced
GPS trajectory using these durations. The result is staggering: only
$11\%$ of the trips are correctly sampled. One may argue that calls
and rests are correlated, or that calls done during moves can be
filtered out. We thus computed the proportion of correctly sampled
trips at different levels of correlation (see SI for details), and
find that, at best---when we only have calls during rests---only
$16\%$ of trips are recovered. The use of CDR mobile phone data or of
any dataset presenting a largely distributed inter-event time to study
mobility is thus very questionable. We note that forcing a perfect
correlation between calls and rests amounts to forcing assumption
$(i)$ presented above. Yet, the trajectory is still poorly sampled,
meaning that assumption $(ii)$ is flawed.

%%%% DISCUSSION %%%%
\section{Discussion}

A key aspect of every experimental science is to be aware of the
limits of the experiment's setup and of the measuring
apparatus. Unfortunately, this point has often been neglected in the
recent trend of data-driven studies. The greed for novel, large-impact
results is leading to studies where many corners are cut. As a
consequence, a large number of quantitative results are sustained
almost exclusively by the sheer amount of data gathered, even when
those data are not adequate for the problem at hand: not all biases do
average themselves out.
This is particularly true for the
study of trajectories from sampling movements in space. The
choices taken for trajectory segmentation, together with the temporal
and spatial granularity of the measures, influence all quantities associated to these trajectories~\cite{Laube:2011}.
%and cast some shadow on the current empirical results.

In this paper, we have shown that for any sampling of a trajectory
alternating rests and movements (of animals, human, or artefacts) the assumptions that each measure correspond to
a rest and that an observed displacement correspond to a move are
intrinsically flawed. The same is true for any sampling of a trajectory alternating
rests and movements (of animals, humans, or artefacts).  We solved
analytically an
idealized %RG: not sure about `idealized', perhaps ideal, ALSO IF WE HAVE ARGUMENT FOR THIS BEING A BEST CASE SCENARIO WE REMARK IT HERE
case which shows that the fraction of trips that are correctly
identified with a constant sampling time interval is \emph{at best}
$51\%$. We also showed that this fraction is significantly lower in
any other realistic scenario, especially when mobility is being
studied through the lens of mobile phone communications: using phone
calls in order to track mobility gives correct predictions for $23\%$
of the trips made with a car. Result get even worse if one wants to
investigate mobility at a finer scale: using high-resolution GPS data
the value drops down to $11\%$, and we estimate that no more of $16\%$
of movements can be recovered, even if a perfect stay-point
identification algorithm is applied. These figures (summarized in
the Table S1) cast a shadow on
the possibility of understanding~\cite{Gonzalez:2008} and
modeling~\cite{Song:2010mod} human mobility from CDR
data. Our ability of predicting individuals
movements~\cite{Song:2010lim} is not only limited by temporal and
spatial scale of analysis~\cite{Gallotti:2013,Cuttone:2016} , but also
and highly predominantly by limitations inherent to the data
sources. Our results highlight the general fact that a
rigorous analysis of the empirical methods used in many studies is
necessary in order to construct solid foundations for our knowledge. We
provided new analytical tools to evaluate the quality of a sampled
trajectory for the study of both animal and human movements. Positions
must be collected at least with a frequency commensurate with the
underlying moving and resting dynamics
($\dbar \approx 1.96\sqrt{\tbar\taubar}$). Alternatively, stay points
can be reconstructed from high-frequency sampling
($\langle \Delta \rangle \ll \taubar$), but not when one has bursty
inter-event times, because during the numerous extreme events
constituting the long tail of the distribution $P(\Delta)$ the
information on the movements is simply absent. Further studies and rigorous analysis of the
empirical methods used in many studies are thus necessary in order to
construct solid foundations for our knowledge.

\section{Methods}

\subsection{GPS data}
%\subsection{The data}
    %\label{sec:The data}

In order to prove the validity of our claims, we test the predictions presented
in the main text on high-resolution data, the GeoLife GPS
trajectories~[54].
%\cite{Geolife3}. 
This dataset consists in the
trajectories of 182 subjects registered by a GPS device over the course of 3
years. The database contains  $17,621$ trajectories for a cumulated travel
length of more than $1,000,000$ km. Most trajectories are logged with a
temporal precision of the order of the second.

Because the term of `rest' has a behavioural connotation, we will talk in the
following about stay points~[40]. These are locations where an individual stays
for a certain period of time and from which she does not depart too much. Of
course, the identification of stay points depends on the spatial and temporal
granularity of the data~[20].

As mentioned in the introduction, the absence of contextual information
forces us to make more or less realistic assumptions in order to identify
traveling times and rests. We begin by filtering out the trajectories that
are less than $1$ km long, as they are not representative. We then proceed to
identify stay points as follows:

\begin{itemize}
    \item We consider all points around the point $p_t$ in a moving time window of duration $\tau = 10$ s around $t$;
    \item In this window, we compute the average movement speed between
        successive trajectory points;
    \item If the average speed is lower than $2 \text{ m/s}$ (fast walker),
        we identify $p_t$ as a stay point;
    \item We iterate the procedure for all points in the trajectory and
        aggregate consecutive stay points.
\end{itemize}

The averaging is introduced in order to minimize the impact of fluctuations in the
GPS reading. After this procedure, we obtain individual trajectories where stay
points are identified.

We find the average travel and rest times $\overline{t} = 20$~min and
$\overline{\tau} = 48$~min. The average travel time is
identical to that observed for vehicular mobility.
The average duration $\taubar$ of a rest is however significantly shorter.

\subsection{CDR data}

We use the dataset 2 `fine-grained mobility' of the Orange data made available for the D4D challenge~[56]
%\cite{DeMontjoye:2014}  
that provides anonymized individual CDR records. For privacy reasons, the caller ids are reshuffled every 15 days. The dataset spans 25 such 15-day periods.
The selection procedure which is most often used is the one proposed in~[35],
i.e., selecting only the individuals whose average call frequency is greater than 0.5 calls/hour. Here, we allowed for a more conservative margin by selecting only the $1.1\%$ of individuals who had more than 1 call/hour in a period of 15 days.
Furthermore, the data provide call time stamps with a 10-minutes granularity.
We apply a smoothing procedure that consists in picking a time uniformly at random between $M-5$ and $M+5$, where $M$ is the value in minutes indicated by the time stamp.
One should bear in mind that the mobile phone CDR and GPS trajectories come from two independent datasets describing two different populations and times of the year. For this reason, we did not enforce calendar synchronization between the datasets, but used the CDR data to randomly extract real inter-event times with the appropriate minimal frequency. More accurate numbers would thus be obtained in a situation where informations on calls and trajectories would be available for the same user.

\subsection{Characteristic times for car mobility}

We need to identify the values of $\tbar$ and $\taubar$ in conditions
that realistically describe human mobility. We do this by using the
results of the analysis of urban and inter-urban traffic of private vehicles
in Italy~[7].

The average travel time observed for Italian cars is
$\tbar = 0.30$~h. Moreover, as discussed in~[7] and references therein,  in
privateas in
public transportation, the distribution of trip
durations $P(t)$ in a city is short-tailed. A similar result has been
found in taxi rides, in survey data
(where also $\tbar\approx 0.30$~h) and on the GPS data~[55]
%\cite{Zhao:2015} 
we use in this
work (for separated modes of transport). For this
reason, we can safely limit our numerical analysis to the case of
exponential $P(t)$.

Concerning rest times, two different functional forms have been
proposed for the distribution $P(\tau)$. Car parking
durations have been fitted with a stretched exponential:
\begin{equation}
P(\tau)= \frac{\exp(-(\tau/\tau_0)^\beta)}{\tau_0\Gamma(1+1/\beta)},
\label{eq:tauSE}
\end{equation}
with $\tau_0\approx 10^{-4}$~h and $\beta\approx 0.19$~[7].
For mobile phone data, a truncated power-law fit has been proposed:
\begin{equation}
P(\tau)\propto \tau^{-\gamma}\exp(-\tau/\tau_e),
\label{eq:tauTPL}
\end{equation}
with $\gamma\approx 1.8$ and $\tau_e\approx 17$~h~[4].
This fit is made on movements sampled at best with $\overline\Delta = 1$~h
(it is thus expected to be influenced by the
sampling issues described in the main text), and does not allow to
identify rests shorter than 1 h. Note that in estimating the
distribution's average below, we are extending the distribution
(\ref{eq:tauTPL}) below this experimental range.

Averaging the distributions (\ref{eq:tauSE}) and (\ref{eq:tauTPL})
between 5 minutes and 24 hours, which corresponds to selecting only
individuals moving every day, we obtain average rest times of 2.49 h
and 0.55 h respectively. To have a consistent description of car
mobility, we choose to use the value $\taubar = 2.49$~h. Since our
results suggest that the larger $\taubar$ the better the sampling, our
choice also defines the best-case scenario.

\begin{acknowledgments}
RG, RL, JML, and MB designed the research. RG
performed the numerical analysis. RL
performed the data analysis and JML performed the analytical
calculations. RG and RL prepared the figures. RG, RL, JML and MB wrote the text. RG thanks M. Lenormand and T. Louail for interesting discussions.
\end{acknowledgments}

%%%%%%%%%%%%%%%%%%%%%%%%%%%%%%%%%%%%%%%%%%%%%%%%%%%%%%%%%%%%
%%%%%%%%%%%%%%%%%%% APPENDIX %%%%%%%%%%%%%%%%%%%%%%
%%%%%%%%%%%%%%%%%%%%%%%%%%%%%%%%%%%%%%%%%%%%%%%%%%%%%%%%%%%%

%\onecolumngrid
\appendix

%\newpage
\section{APPENDIX: Analytical calculations}

%\rg{Please mind that the sampled distances are $\ell^\ast$ while the
%  original distance are $\ell$}

\subsection{Possible sampling scenarios}

Sampling can get wrong in seven different ways. The cases are the following. We can have two sampling times falling:

\begin{itemize}
    \item[$1.$] in the same rest;
    \item[$2a.$] in two subsequent rests (the correct way);
    \item[$2b.$] in two rests separated by more than one move;
    \item[$3a.$] in a move and in the rest following that move;
    \item[$3b.$] in a move and in a rest not following that move;
    \item[$4.$] in the same move;
    \item[$5a.$] in a rest and in the move following that rest;
    \item[$5b.$] in a rest and in a move not following that rest;
    \item[$6a.$] in two subsequent moves;
    \item[$6b.$] in two non-subsequent moves.
\end{itemize}

Case $1$ can be identified, since the displacement is $\ell=0$. Case $2a$ gives a correct evaluation of the move performed, since both sampling are made when the individual is still and only one movement has been done in that time.
Cases $3a$, $4$, $5a$ and $6a$ `cut' moves, under-estimating the observed displacements and leading to an over-estimate of the number of moves.
Cases $2b$, $3b$, $5b$ and $6b$ `join' together multiple moves, thus yielding over-estimated displacements and under-estimated number of moves.

\subsection{General setting}

The random trajectory
consists in an alternation of moves with durations $t_1, t_2, t_3,\dots$,
where the position $x(\theta)$ increases with unit velocity ($v=1$),
and of rests with durations $\tau_1,\tau_2,\tau_3,\dots$,
where $x(\theta)$ stays constant.
The walker starts from $x=0$ at time $\theta=0$.
The move durations $t_k$ and the rest durations $\tau_k$
are drawn from two given continuous distributions $f(t)$ and $g(\tau)$.

We are interested in the distribution $P_{\theta_1,\theta_2}(\ell)$ of the distance
\beq
\ell=x(\theta_2)-x(\theta_1)
\eeq
traveled by the walker between two fixed times $\theta_1$ and $\theta_2$,
and in various related quantities.
An exact expression for the distribution $P_{\theta_1,\theta_2}(\ell)$
can be derived by analytical means,
for arbitrary distributions $f(t)$ and $g(\tau)$,
by using techniques from renewal theory~[45,46,47].
%\cite{feller,cox,coxm}.
The key quantity is the triple Laplace transform
\beqa
\L(r,s,u)\esp&=&\esp
\int_0^\infty\e^{-r\theta_1}\,\dd\theta_1
\int_{\theta_1}^\infty\e^{-s\theta_2}\,\dd\theta_2
\nonumber\\
\esp&\times&\esp\int_0^\infty\e^{-u\ell}\,P_{\theta_1,\theta_2}(\ell)\,\dd\ell.
\eeqa

\begin{figure}[ht!]
\begin{center}
\includegraphics[angle=0,width=0.45\textwidth]{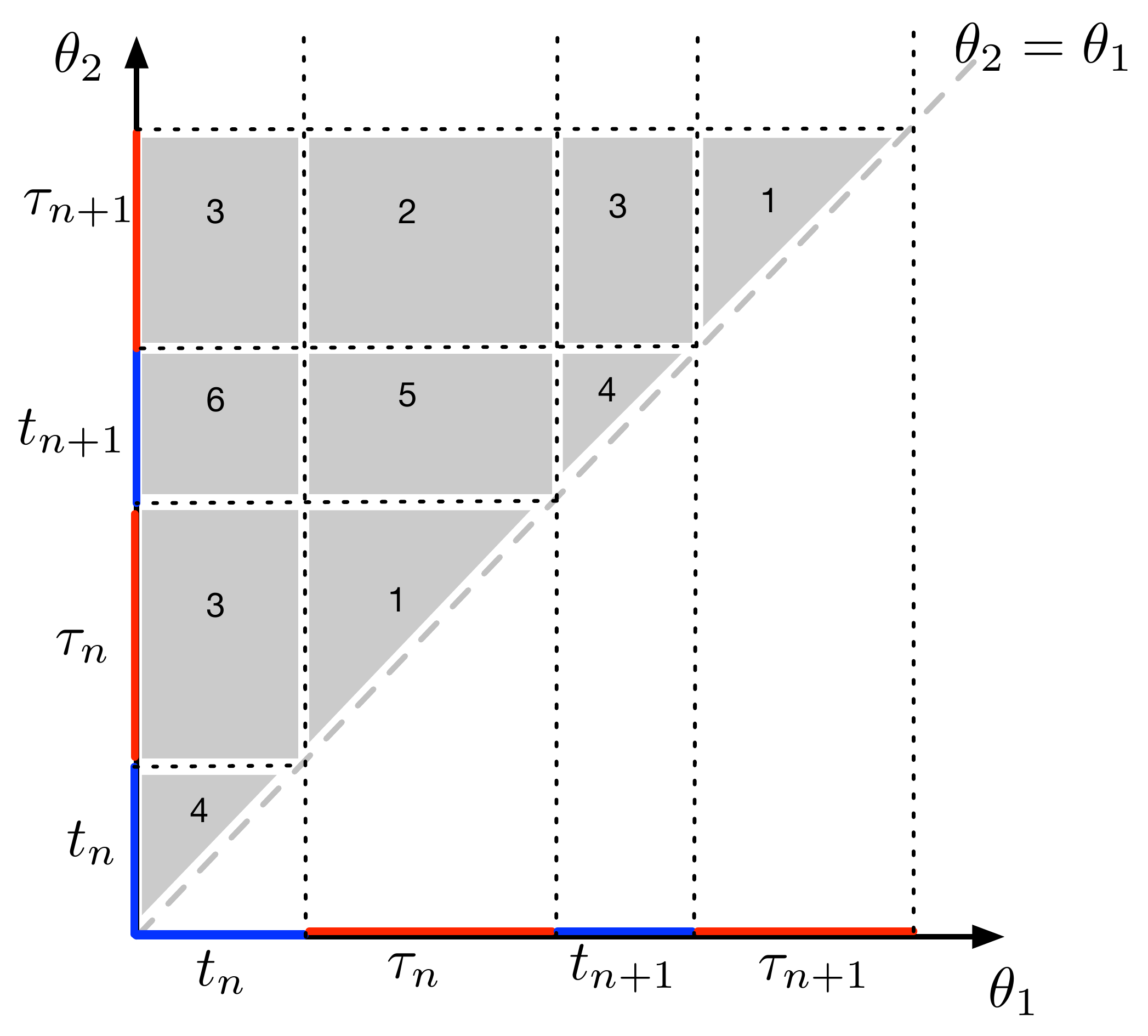}
\end{center}
\caption{{\bf Schematic representation of the different sectors.} In
  the plane $(\theta_1,\theta_2)$ we represent the different sectors
  according to the numbering defined in the text. }
\label{fig:sectors}
\end{figure}

This quantity can be evaluated as a sum over six sectors
(see above discussion and Fig.~\ref{fig:sectors}):
\begin{itemize}
\item[$1.$] $\theta_1$ and $\theta_2$ belong to the same $\tau_n$;
\item[$2.$] $\theta_1$ belongs to $\tau_m$ while $\theta_2$ belongs to $\tau_n$;
\item[$3.$] $\theta_1$ belongs to $t_m$ while $\theta_2$ belongs to $\tau_n$;
\item[$4.$] $\theta_1$ and $\theta_2$ belong to the same $t_n$;
\item[$5.$] $\theta_1$ belongs to $\tau_m$ while $\theta_2$ belongs to $t_n$;
\item[$6.$] $\theta_1$ belongs to $t_m$ while $\theta_2$ belongs to $t_n$.
\end{itemize}

Let us illustrate the method on the example of sector~2.
For fixed integers $m\ge1$ and $n\ge m+1$, we have
\beqa
\theta_1\esp&=&\esp\Theta_1+B_1,
\nonumber\\
\theta_2\esp&=&\esp\Theta_2+B_2,
\nonumber\\
\Theta_1\esp&=&\esp(t_1+\cdots+t_m)+(\tau_1+\cdots+\tau_{m-1}),
\nonumber\\
\Theta_2\esp&=&\esp(t_1+\cdots+t_n)+(\tau_1+\cdots+\tau_{n-1}),
\nonumber\\
x(\theta_1)\esp&=&\esp t_1+\cdots+t_m,
\nonumber\\
x(\theta_2)\esp&=&\esp t_1+\cdots+t_n,
\nonumber\\
\ell\esp&=&\esp t_{m+1}+\cdots+t_n,
\eeqa
with $0<B_1<\tau_m$ and $0<B_2<\tau_n$.
The contribution of sector~2 with fixed $m$ and $n$
to $\L(r,s,u)$ therefore reads
\beqa
\L_2^{(m,n)}(r,s,u)\esp&=&\esp\biggl\langle\e^{-r\Theta_1-s\Theta_2-u\ell}
\nonumber\\
\esp&\times&\esp\int_0^{\tau_m}\!\e^{-rB_1}\,\dd B_1
\int_0^{\tau_n}\!\e^{-sB_2}\,\dd B_2\biggr\rangle.
\label{l2def}
\eeqa
Hereafter we borrow conventions and notations from Ref.~[41].
%\cite{Godreche:2001}.
In particular, $\mean{\dots}$ denotes an average over the random process,
i.e., over all the move durations $t_k$ and rest durations~$\tau_k$.

The explicit evaluation of~(\ref{l2def}) involves three steps.

\noindent $\bullet$ First, performing the two integrals leads to the expression
\beq
\L_2^{(m,n)}(r,s,u)=\bigmean{\e^{-r\Theta_1-s\Theta_2-u\ell}
\frac{1-\e^{-r\tau_m}}{r}\frac{1-\e^{-s\tau_n}}{s}},
\eeq
which only involves the $t_k$ and $\tau_k$.

\noindent $\bullet$ Second, averaging independently over all the $t_k$ and $\tau_k$,
we obtain
\beqa
\L_2^{(m,n)}(r,s,u)\esp&=&\esp\hf(r+s)^m\hf(s+u)^{n-m}\hg(r+s)^{m-1}
\nonumber\\
\esp&\times&\esp\hg(s)^{n-m-1}\,\frac{\hg(s)-\hg(r+s)}{r}\,\frac{1-\hg(s)}{s}
\eeqa
in terms of the Laplace transforms (characteristic functions)
\beqa
\hf(s)\esp&=&\esp\mean{\e^{-st}}=\int_0^\infty f(t)\,\e^{-st}\,\dd t,
\nonumber\\
\hg(s)\esp&=&\esp\mean{\e^{-s\tau}}=\int_0^\infty g(\tau)\,\e^{-s\tau}\,\dd\tau.
\eeqa

\noindent $\bullet$ Third, the entire contribution of sector~2 reads
\beq
\L_2(r,s,u)=\sum_{m=1}^\infty\sum_{n=m+1}^\infty\L_2^{(m,n)}(r,s,u),
\eeq
where the sums boil down to geometric sums.
This leads to
\beqa
\L_2(r,s,u)\esp&=&\esp\frac{\hf(s+u)}{1-\hf(s+u)\hg(s)}
\,\frac{\hf(r+s)}{1-\hf(r+s)\hg(r+s)}
\nonumber\\
\esp&\times&\esp\frac{\hg(s)-\hg(r+s)}{r}\,\frac{1-\hg(s)}{s}.
\eeqa

The contributions of the five other sectors can be evaluated along the same lines.
We thus obtain
\beqa
\L_1(r,s,u)\esp&=&\esp\frac{\hf(r+s)}{1-\hf(r+s)\hg(r+s)}
\nonumber\\
\esp&\times&\esp\frac{r+s\hg(r+s)-(r+s)\hg(s)}{rs(r+s)},
\nonumber\\
\L_3(r,s,u)\esp&=&\esp\frac{1}{1-\hf(s+u)\hg(s)}\,\frac{1}{1-\hf(r+s)\hg(r+s)}
\nonumber\\
\esp&\times&\esp\frac{\hf(s+u)-\hf(r+s)}{r-u}\,\frac{1-\hg(s)}{s},
\nonumber\\
\L_4(r,s,u)\esp&=&\esp\frac{1}{1-\hf(r+s)\hg(r+s)}
\nonumber\\
\esp&\times&\esp\frac{u-r+(r+s)\hf(s+u)-(s+u)\hf(r+s)}{(r+s)(u-r)(s+u)},
\nonumber\\
\L_5(r,s,u)\esp&=&\esp\frac{1}{1-\hf(s+u)\hg(s)}\,\frac{\hf(r+s)}{1-\hf(r+s)\hg(r+s)}
\nonumber\\
\esp&\times&\esp\frac{\hg(s)-\hg(r+s)}{r}\,\frac{1-\hf(s+u)}{s+u},
\nonumber\\
\L_6(r,s,u)\esp&=&\esp\frac{\hg(s)}{1-\hf(s+u)\hg(s)}\,\frac{1}{1-\hf(r+s)\hg(r+s)}
\nonumber\\
\esp&\times&\esp\frac{\hf(r+s)-\hf(s+u)}{u-r}\,\frac{1-\hf(s+u)}{s+u}.
\eeqa

\subsection{Steady state}

From now on we focus our attention onto the steady state of the process,
obtained by letting the first time $\theta_1$ go to infinity, keeping the time difference
\beq
\del=\theta_2-\theta_1
\eeq
fixed.
This steady state is well-defined if the distributions $f(t)$ and $g(\tau)$
decay fast enough for the mean values $\tbar$ and $\taubar$ to be finite.
In the opposite situation, where either $\tbar$ or $\taubar$ or even both are divergent,
the process never reaches a steady state.
It rather exhibits various non-stationary features,
usually referred to as aging or weak ergodicity breaking~[48].
%\cite{bel,metzler,schultz}.
We assume henceforth that $\tbar$ and $\taubar$ are finite.

The quantity of most interest is the steady-state distribution $P_\del(\ell)$.
Its double Laplace transform
\beq
L(s,u)=
\int_0^\infty\e^{-s\del}\,\dd\del
\int_0^\infty\e^{-u\ell}\,P_\del(\ell)\,\dd\ell
\eeq
is the limit of the product $(r+s)\L(r,s,u)$ as $r\to-s$.
We thus obtain
\beq
L(s,u)=\frac{N(s,u)}{(\tbar+\taubar)s^2(s+u)^2(1-\hf(s+u)\hg(s))},
\label{lsu}
\eeq
with
\beqa
N(s,u)\esp&=&\esp s(s+u)(s\tbar+(s+u)\taubar)(1-\hf(s+u)\hg(s))
\nonumber\\
\esp&-&\esp u^2(1-\hf(s+u))(1-\hg(s)).
\label{nsu}
\eeqa

The distribution $P_\del(\ell)$ has the general form
\beq
P_\del(\ell)=C_0(\del)\,\delta(\ell)+C_1(\del)\,\delta(\ell-\del)
+P_\del^\cont(\ell),
\label{pgal}
\eeq
with two delta functions at the endpoints $\ell=0$ and $\ell=\del$,
and a non-trivial continuous piece in-between.
The delta function at $\ell=0$ corresponds to sector~1
($\theta_1$ and $\theta_2$ belong to the same rest duration $\tau_n$).
The Laplace transform $\h C_0(s)$ of the associated amplitude $C_0(\del)$ reads
\beq
\h C_0(s)=\frac{\hg(s)+s\taubar-1}{(\tbar+\taubar)s^2},
\label{c0lap}
\eeq
hence
\beq
C_0(\del)=\frac{1}{\tbar+\taubar}\int_\del^\infty(\tau-\del)g(\tau)\,\dd\tau.
\eeq
Similarly, the delta function of $P_\del(\ell)$ at $\ell=\del$ corresponds to sector~4
($\theta_1$ and $\theta_2$ belong to the same move duration $t_n$).
The associated amplitude reads
\beq
C_1(\del)=\frac{1}{\tbar+\taubar}\int_\del^\infty(t-\del)f(t)\,\dd t.
\eeq
As $\del$ increases from 0 to infinity,
the above amplitudes decrease monotonically
from $C_0(0)=\taubar/(\tbar+\taubar)$ and $C_1(0)=\tbar/(\tbar+\taubar)$ to zero,
whereas the weight of the continuous part $P_\del^\cont(\ell)$
increases from zero to one.

The mean value of $\ell$ has the remarkably simple expression
\beq
\mean{\ell}_\del=\frac{\tbar}{\tbar+\taubar}\,\del.
\label{ave}
\eeq
The second moment of $\ell$ reads in Laplace space
\beqa
\int_0^\infty\e^{-s\del}\mean{\ell^2}_\del\,\dd\del
\esp&=&\esp\frac{2\tbar}{(\tbar+\taubar)s^3}
\nonumber\\
\esp&-&\esp\frac{2(1-\hf(s))(1-\hg(s))}{(\tbar+\taubar)s^4(1-\hf(s)\hg(s))}.
\eeqa
This formula can be inverted in the regime where $\del$ is much larger
than $\tbar$ and $\taubar$, yielding
\beq
\mean{\ell^2}_\del-\mean{\ell}_\del^2\approx
\frac{(\tauq-\taubar^2)\tbar^2+(\tq-\tbar^2)\taubar^2}{(\tbar+\taubar)^3}\,\del+K.
\eeq
The linear growth of the variance of $\ell$ with $\del$
testifies that the continuous part $P_\del^\cont(\ell)$
satisfies an approximate central limit theorem at large $\del$,
where the measured displacement $\ell$ is the sum
of a typically large number of elementary moves.
The constant $K$, corresponding to the first correction to this limit,
is given by a combination of moments of $t$ and $\tau$,
which can be either positive or negative.

Another quantity which is used in the main text
is the fraction $P_\good(\del)$ of the moves which are correctly sampled.
These events correspond to the observation times $\theta_1$ and $\theta_2$
belonging to two consecutive rests surrounding the move under consideration,
i.e., to sector~2 with $n=m+1$.
It is also useful to introduce the normalised fraction of correctly sampled moves,
\beq
F_\good(\del)=\frac{P_\good(\del)}{1-C_0(\del)},
\eeq
where the denominator is nothing but the probability of measuring a non-zero displacement.

In the steady-state of the process, we obtain in Laplace space
\beq
\h P_\good(s)=\frac{\hf(s)(1-\hg(s))^2}{(\tbar+\taubar)s^2},
\label{fg}
\eeq
whereas the Laplace transform of $C_0(\del)$ is given by~(\ref{c0lap}).

\subsection{Exponential distributions}

When the distributions of the move and rest durations
are exponential, with respective parameters $a=1/\tbar$ and $b=1/\taubar$, i.e.,
$f(t)=a\,\e^{-at}$,
$g(\tau)=b\,\e^{-b\tau}$,
$\hf(s)=a/(s+a)$,
$\hg(s)=b/(s+b)$,
the above expressions simplify drastically,
and so many observables can be evaluated in closed form.

Eqs.~(\ref{lsu}),~(\ref{nsu}) read
\beq
L(s,u)=\frac{(a+b)^2+(a+b)s+au}{(a+b)(s^2+(a+b+u)s+bu)}.
\label{lexp}
\eeq

We thus recover~(\ref{ave}), i.e.,
\beq
\mean{\ell}_\del=\frac{b}{a+b}\,\del,
\eeq
as well as the following expression
\beq
\mean{\ell^2}_\del=\mean{\ell}_\del^2+\frac{2ab}{(a+b)^3}\,\del
+\frac{2ab}{(a+b)^4}\left(\e^{-(a+b)\del}-1\right)
\eeq
for the second moment of the measured displacement.
In this example, the constant $K=-2ab/(a+b)^4$ is negative.
Higher moments can be evaluated as well.

With the notations of the main text,
i.e., in terms of $\taubar,\tbar,\ell^\ast,v,\dbar$, the first two moments read
\begin{equation}
\langle \ell^\ast \rangle = \frac{v\dbar}{1+\taubar/\tbar}
\label{eq:meanTeo}
\end{equation}
and
\begin{align}
\nonumber
\langle \ell^{\ast 2} \rangle &= \frac{v^2\dbar^2}{(1+\taubar/\tbar)^2} +
  \frac{2v^2\taubar^2\dbar}{\tbar(1+\taubar/\tbar)^3} \\
%&+  \frac{2\taubar^3}{\tbar(1+\taubar/\tbar)^4}\left(\exp\left({-\frac{(\tbar+\taubar)\dbar}{\tbar\taubar}}\right)-1\right) \,.
&+  \frac{2v^2\taubar^3}{\tbar(1+\taubar/\tbar)^4}\left(\exp\left({-\frac{(\tbar+\taubar)\dbar}{\tbar\taubar}}\right)-1\right).
\end{align}

The full distribution $P_\del(\ell)$ can also be obtained in closed form.
In a first step,
performing the inverse transform of the expression~(\ref{lexp}) from $s$ to $\del$,
we obtain an expression for the Laplace transform
\beq
L_\del(u)=\int_0^\infty\e^{-u\ell} P_\del(\ell)\,\dd\ell,
\eeq
namely
\beqa
L_\del(u)\esp&=&\esp\e^{-(a+b+u)\del/2}
\nonumber\\
\esp&\times&\esp\!\left(\!\cosh R+\frac{(a+b)^2+(a-b)u}{2(a+b)}\frac{\sinh R}{R}\del\!\right)\!,
\label{lu}
\eeqa
with
\beq
R=\frac{\del}{2}\,\sqrt{(a-b+u)^2+4ab}.
\eeq
In a second step, the expression~(\ref{lu}) can be inverse transformed from $u$ to $\ell$,
yielding an end result of the expected general form~(\ref{pgal}),
with
\beq
C_0(\del)=\frac{a}{a+b}\,\e^{-b\del},\quad
C_1(\del)=\frac{b}{a+b}\,\e^{-a\del}
\eeq
and
\beqa
P_\del^\cont(\ell)\esp&=&\esp\frac{2ab}{a+b}\,\e^{-a\ell-b(\del-\ell)}
\nonumber\\
\esp&\times&\esp
\left(I_0(x)+(a(\del-\ell)+b\ell)\frac{I_1(x)}{x}\right),
\eeqa
with
\beq
x=2\sqrt{ab\ell(\del-\ell)},
\eeq
and where $I_0$ and $I_1$ are modified Bessel functions.

The expression~(\ref{fg}) reads
\beq
\h P_\good(s)=\frac{a^2b}{(a+b)(a+s)(b+s)^2}.
\eeq
We have therefore
\beqa
P_\good(\del)\esp&=&\esp\frac{a^2b}{(a+b)(a-b)^2}
\nonumber
\\
\esp&\times&\esp\left(\e^{-a\del}+((a-b)\del-1)\e^{-b\del}\right)
\eeqa
and
\beq
F_\good(\del)=\frac{a^2b}{(a-b)^2}\,\frac{\e^{-a\del}+((a-b)\del-1)\e^{-b\del}}{a+b-a\,\e^{-b\del}}.
\label{fgexp}
\eeq

%\rg{I guess here the quantities that are with `hats' in the main
%  text are with asterisk here because hats denote Laplace
%  transform. Should we point it out with one sentence (or considering
%  renaming all hatted quantities in figures and main text at a second
%  submission?})

The normalised fraction $F_\good(\del)$ of correctly sampled moves
starts growing as $(a\del)^2/2$ at small $\del$,
whereas it falls off exponentially at large $\del$.
It therefore reaches a non-trivial maximum $\hat F_\good$
for an optimal value $\hat\del$ of $\del$ (see Fig.~S2).
In the limit $b/a\to0$,
the maximal value $\hat F_\good$ reaches unity.
It is attained for
\beq
\hat\del\approx\frac{2}{\sqrt{ab}}.
\eeq
It however drops from this perfect value very rapidly,
with a square-root singularity of the form
\beq
\hat F_\good\approx1-2\sqrt\frac{b}{a}.
\eeq
For $a=b$, the expression~(\ref{fgexp}) simplifies to
\beq
F_\good(\del)=\frac{(a\del)^2}{2(2\e^{a\del}-1)}.
\eeq
The maximal value is already as small as $\hat F_\good\approx0.14602$.
It is reached for $\hat\del\approx1.84141/a$.

%%%%%%%%%%%%%%%%%%%%%%%%%%%%%%%%%%%%%%%%%%%%%%%%%%%%%%%%%%%%
%%%%%%%%%%%%%%%%%%% SUPPLEMENTARY INFORMATION %%%%%%%%%%%%%%%%%%%%%%
%%%%%%%%%%%%%%%%%%%%%%%%%%%%%%%%%%%%%%%%%%%%%%%%%%%%%%%%%%%%
\clearpage
\onecolumngrid
\renewcommand\thefigure{S\arabic{figure}}
\setcounter{figure}{0}
\renewcommand\thesection{S\arabic{section}}
\setcounter{section}{0}
\newpage

\begin{center}\Huge
 Supplementary Information
\end{center}
\pagestyle{plain}
\setcounter{page}{1}

\section{Supplementary Figures}
%\rg{Figures mentioned in the main text}

\begin{table*}[h!]
\centering
\begin{tabular*}{\textwidth}{@{\extracolsep{\fill}}lcclllrcccccc@{}}
\toprule[1pt]
Mobility represented 	& $\tbar$	(h)	& $\taubar$ (h)	& $P(\tau)$ 				& Sampling		& Type of result 	& $F_{good}$	&	 Notes\\
\bottomrule
Italian cars		& 0.30 		&	2.49		& Exponential				& Periodic 		& Analytical		&	51\%		&	$\dbar = 1.7$h		\\
				&			&			& Stretched exponential 		& Periodic			& Numerical		&	39\%		&	$\dbar = 1.3$h		\\
		 		& 			&			& Exponential		 		& Power law		& Numerical		&	27\%		&		--			\\
		 		& 			&			& Stretched exponential		& Power law		& Numerical		&	23\%		&		--			\\
\hline
USA CDR data 		& (0.30)		& 	0.55		& Exponential				& Periodic			& Analytical		&	24\%		&	$\dbar = 0.8$h		\\
		 		& 			&			& Truncated power law		& Periodic			& Numerical		&	15\%		&	$\dbar = 0.5$h		\\
		 		& 			&			& Truncated power law		& Power law		& Numerical		&	6\%		&		--			\\
\hline
Chinese Geolife Traj. & 0.33		& 	0.80 		& Exponential				& Periodic			& Analytical		&	27\%		&	$\dbar = 1.0$h		\\
				 & 			& 			& Empirical				& Periodic			& Numerical		&	18\%		&	$\dbar = 1.0$h	\\
				 & 			&			& Empirical				& Empirical ($p=0$)	& Numerical		&	11\%		&		--			\\
				 &  			&			& Empirical				& Empirical ($p=1$)	& Numerical		&	16\%		&		--			\\
\bottomrule
\end{tabular*}
\caption{Results' summary.}
\label{table1}
\end{table*}

%\subsection{Representing trajectories}

%%%%%%
\begin{figure*}[ht!]
\begin{center}
\begin{tabular}{cc}
\includegraphics[angle=0,width=0.75\textwidth]{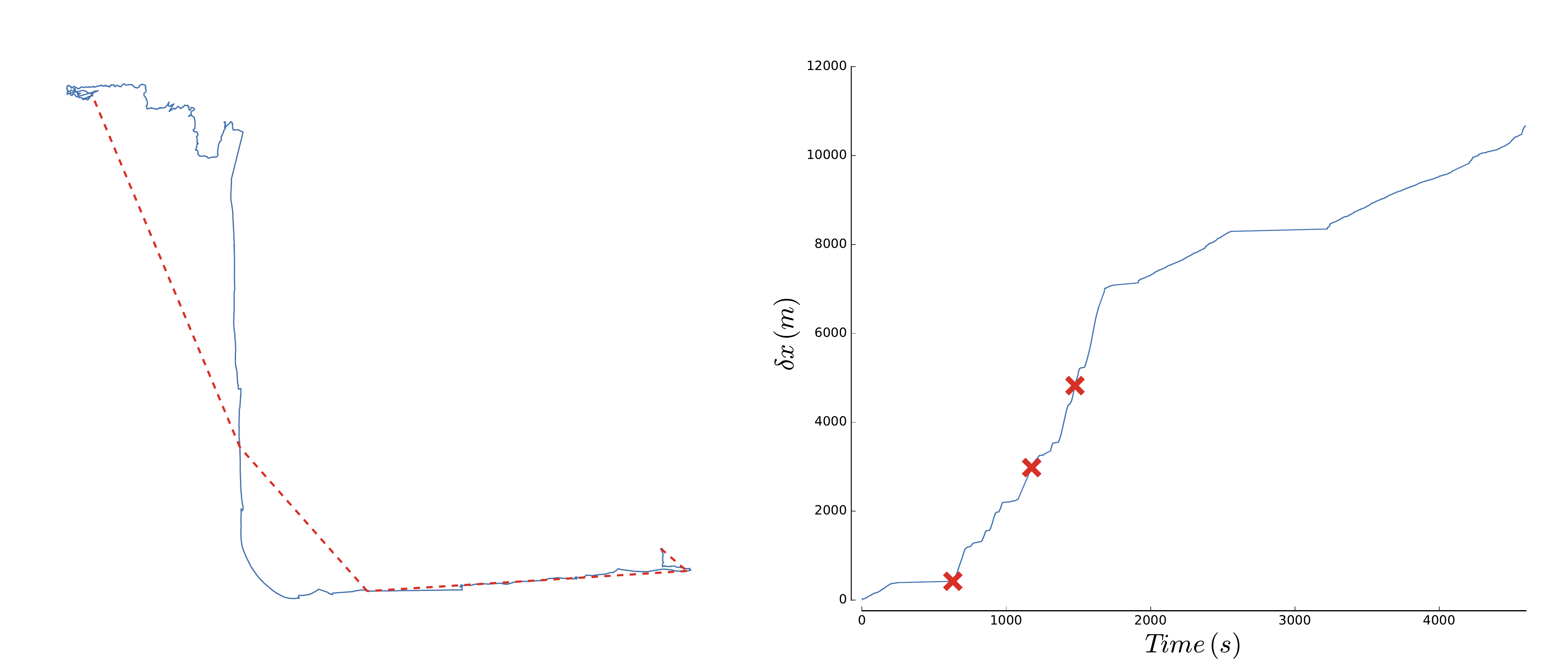}
\end{tabular}
\end{center}
    \caption{{\bf Representing trajectories.} {\bf (Left)} Trajectory obtained with
    GPS readings (solid blue) and sampled trajectory (dashed red). {\bf (Right)}
    Cumulated distance traveled on the real trajectory (solid blue) and
    sampling points (red crosses) drawn from a power-law distribution with
    exponent $-1$. The sampled trajectory is a gross approximation of the real
    trajectory, a lot of information being lost in the
    process.\label{fig:trajectories}
%\rg{Remi please correct this: time is $\theta$, space is $x$, units
%  hours and km, put crosses in the left panel and perhaps make the
%  lines slightly thicker.}
}
\end{figure*}

%\subsection{Moment measured on GPS data}

\begin{figure*}[ht!]
\begin{center}
%\begin{tabular}{cc}
\includegraphics[angle=0,width=0.45\textwidth]{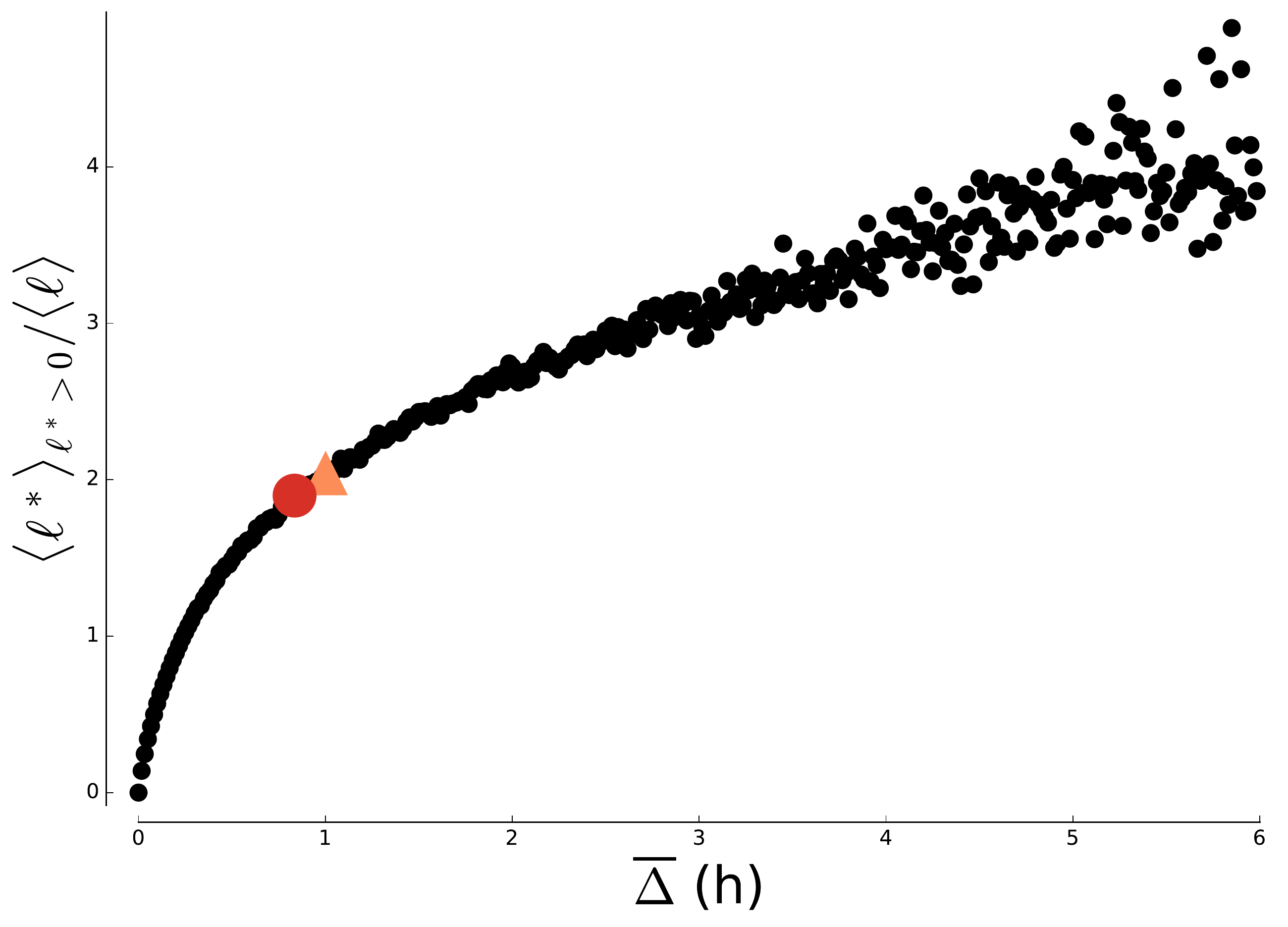}
%\end{tabular}
\end{center}
\caption{{\bf Constant sampling on GPS data.} We present results
  obtained by sampling the GeoLife GPS data with a constant sampling
  interval $\Delta$. We plot the average sampled move
  displacement (computed in 2 dimensions) ($\langle\ell^\ast\rangle_{\ell^\ast >0}$ normalized by the
  real average move length $\langle\ell\rangle$ as a function of the length of the
  sampling interval $\overline{\Delta}$. The ideal case
  $\langle\ell^\ast\rangle_{\ell^\ast>0} = \langle\ell\rangle$
  (Eq.~(9)) is reached for $\dbar=15$~min, while for a short-tailed rest time it is expected to be $\dbar = 52$~min (red circle). The fact that the sampling time optimizing $F_{\rm good}$ (orange triangle) corresponds to $\langle\ell^\ast\rangle_{\ell^\ast>0}/\langle\ell\rangle\approx2$ implies that optimal sampling frequencies would represent, in this case, an under-sampling of the trajectory where moves are more frequently joined together than cut by sampling times.}
\end{figure*}

%\subsection{Rescaled distributions}

%%%%%%
\begin{figure*}[ht!]
\begin{center}
\begin{tabular}{cc}
\includegraphics[angle=0,width=0.45\textwidth]{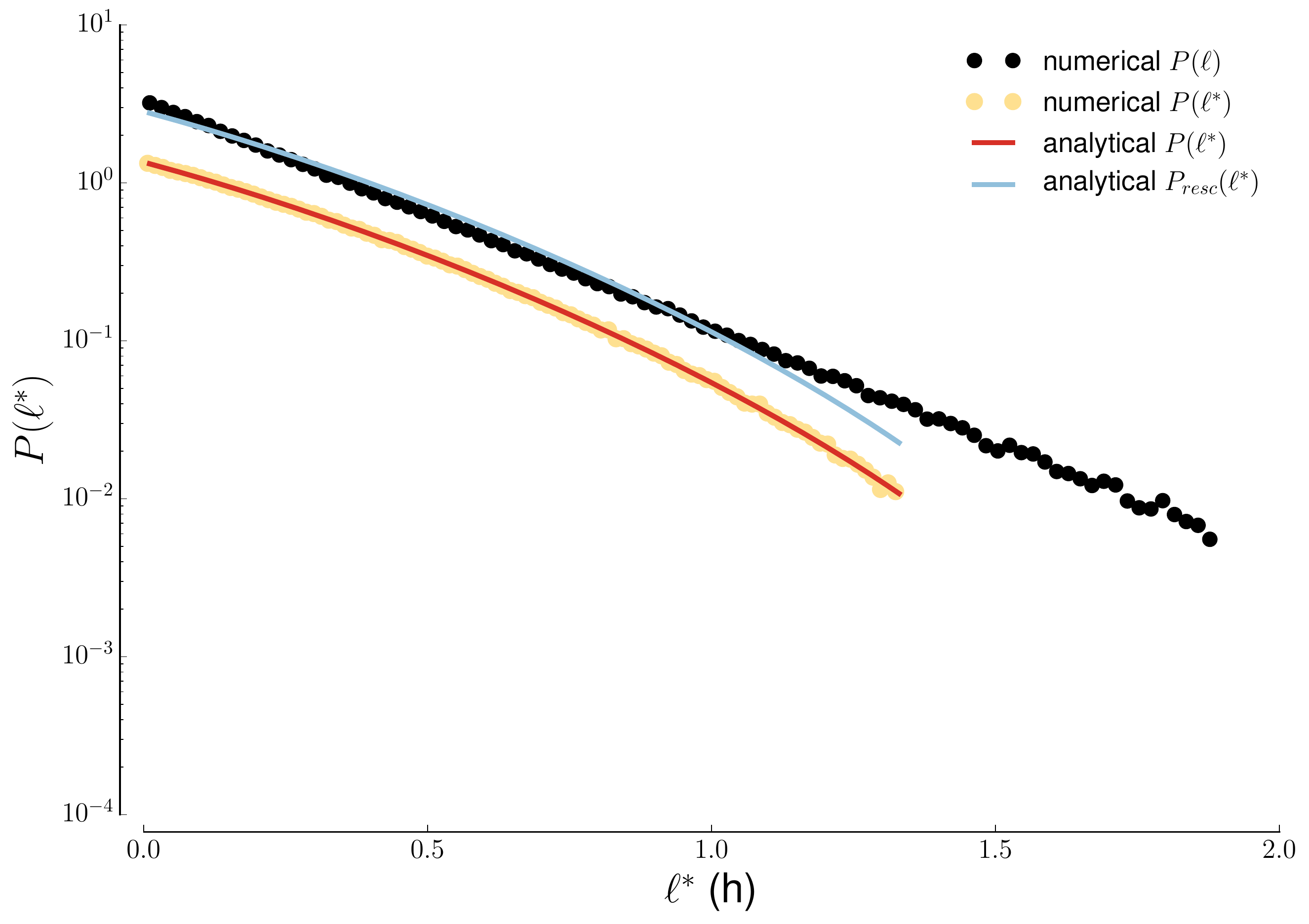}
\end{tabular}
\end{center}
\caption{ {\bf Rescaled distribution of travel times.}
 Optimal Sampling of $\tbar = 0.30$~h with $\dbar = \hat \Delta = 1.73$~h. The distribution of sampled
  travel times (light blue) can be compared with the original exponential
  distribution (black dots) after re-normalizing the distribution, multiplying it
  by the factor $(1-C_0)^{-1}$. }
\label{P_t_resc}
\end{figure*}

%%%%%%%%%%%%%%
\clearpage
\section{Numerical Analysis}
%\rg{I rewrote extensively this section, please check it}

\subsection{Random sampling and long-tailed pause distributions}

\begin{figure*}[ht!]
\begin{center}
\begin{tabular}{cc}
\includegraphics[angle=0,width=0.45\textwidth]{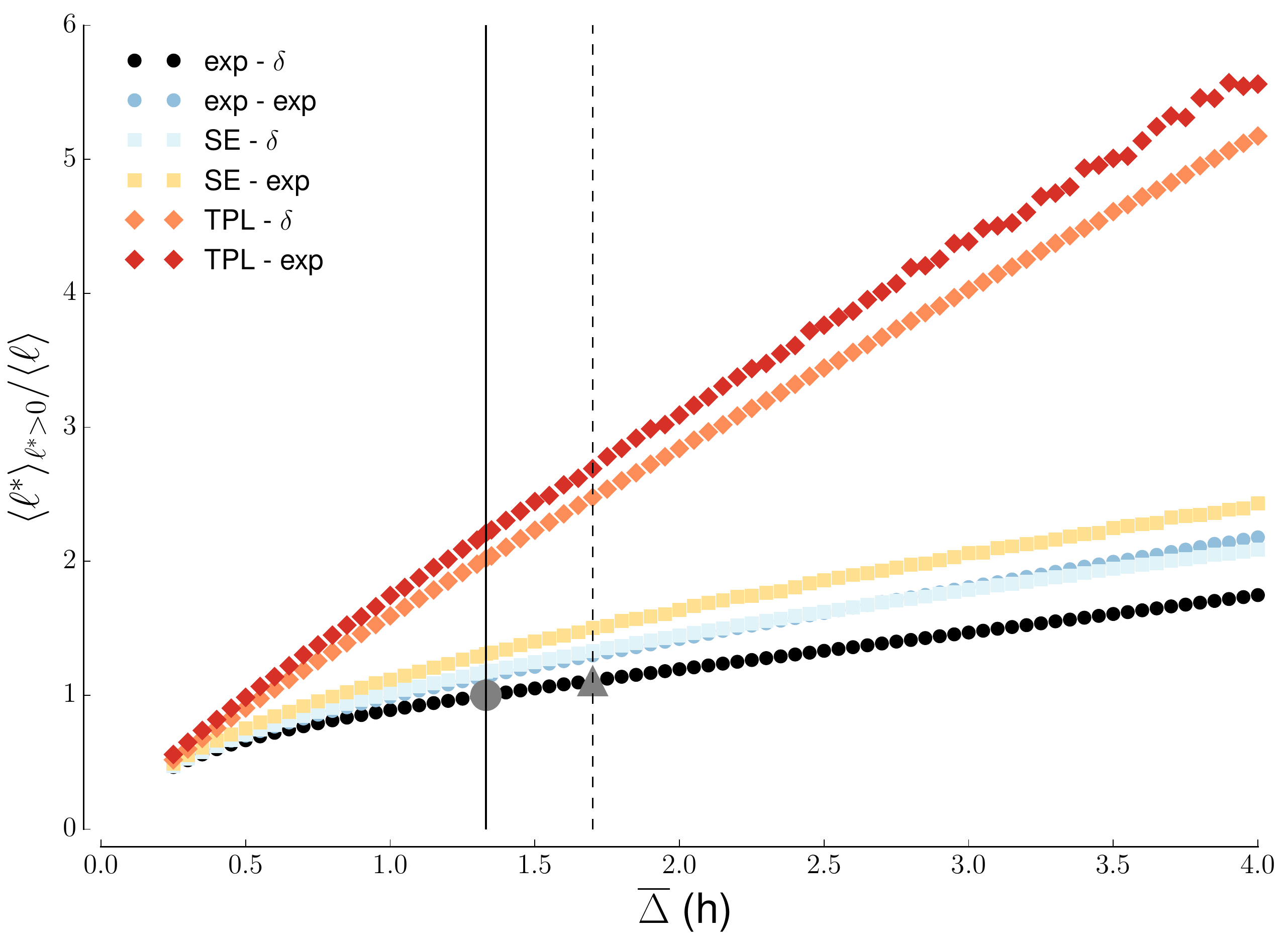}&
\includegraphics[angle=0,width=0.45\textwidth]{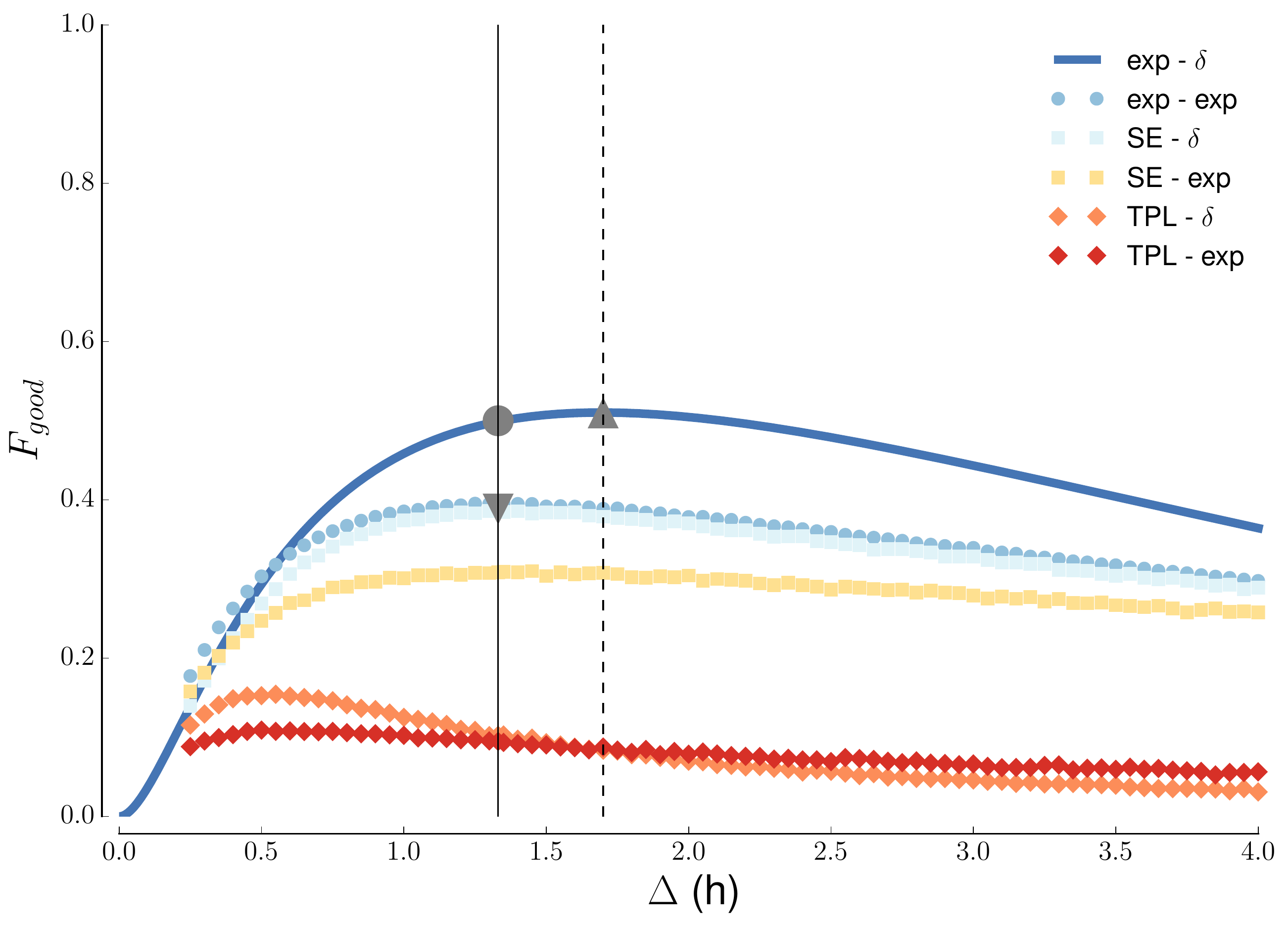}\\
\end{tabular}
\end{center}
\caption{
{\bf Effect of sampling (periodic and Poisson process) with different rest distribution.}
Similar to what we depicted in Fig.~3
(represented here as black circles in the left panel
and as a blue line in the right panel).
In all scenarios studied, we fix $\tbar = 0.30$~h
and consider different alternative forms for $P(\tau)$
(exponential (exp--) or  Stretched Exponential (SE--) or Truncated Power Law (TPL--)).
We use either constant (--$\delta$) or exponentially distributed (--exp) sampling times.
The grey circle represents the value associated with $\mathring\Delta$
and the grey triangle that of $\hat\Delta$ for the (exp -- $\delta$) scenario.
{\bf (Left)} We compare the average value $\langle \ell^\ast\rangle_{\ell^\ast > 0}$ with $\langle \ell \rangle = \overline\ell=v \tbar$.
All definitions of the optimal sampling time which are different from $\mathring\Delta$
(defined by imposing the identity between $\langle \ell^\ast\rangle$ and $\overline\ell$)
are necessarily associated to values smaller than
$\mathring\Delta$ for long-tailed rest distributions.
For the SE case, a sampling time of $\approx 1$h correctly estimates the average.
If rests are distributed as a TPL, the large fraction of short rests
leads to the concatenation of subsequent trips,
even with this relatively short sampling time.
This result illustrates a first incongruence in the work by Song et
al.~[4],
where individual moves and rests are reconstructed with a sampling rate of 1 h,
identifying a rest distribution (the TPL studied here)  for which we predict that only half of the movements would be captured (because $\langle \ell^\ast\rangle/\overline\ell \approx 2$ and consequently $n^\ast/n \approx 1/2$).
 {\bf (Right)}
In all scenarios the fraction $F_{\rm good}$ of correctly sampled trips
is lower than the value (51\%) given by Eq.~(14) and studied extensively in this paper.
In particular, trajectories with long rest times (SE-- and TPL--)
yield worse results than the peaked case (exp--).
At the same time, using exponentially distributed sampling times (--exp)
systematically yields worse results than constant sampling (--$\delta$).
The optimal value $\hat\Delta$ (up triangle and dashed line)
over-estimates the position of the peak for long-tailed rests.
A possible realistic scenario for human mobility is represented by a down triangle,
where trajectories with SE rest distribution
reach a maximum $F_{\rm good}$ of 39\% when sampled
with a constant $\dbar = \check \Delta = 1.33$~h (80 min).
If the pause distribution is TPL, $F_{\rm good}$ barely reaches the $15\%$ mark,
illustrating a second incongruence in the work by Song et al.~[4].}
 \label{fgood_scenarios_1}
\end{figure*}

As stated in the main text, real sampling problems can be more complex than the idealized case defined by Eqs.~(1), (2) and (3). In particular: (i) the rest time distribution can be broad; (ii) sampling times can be random variables; (iii) speed can be a random variable
(Travel durations have been seen to have a short-tailed distribution).
We start by studying the effects of points (i) and (ii),
while we discuss point (iii) in the following section.

Available data on rest durations
suggest that their distribution is long-tailed. Different fits have been
proposed for the latter distribution. Here we consider a Truncated Power Law (TPL),
used for interpreting mobile phone data~[4], and a Stretched
Exponential (SE), used for interpreting private vehicles' parking
times~[7]. As for the sampling time, we can introduce
randomness with an exponential distribution of inter-event times (Poisson
process). Alternatively, if we want to represent the sampling process associated
to communication, we use a power-law distribution with exponent
$-1$~[33,3]. Since this distribution is not
integrable, it is necessarily defined on an interval between
some $\Delta_{\rm min}$ and $\Delta_{\rm max}$.

%%%% FIGURE %%%%
\begin{figure*}[ht!]
\begin{center}
\begin{tabular}{cc}
\raisebox{2.3cm}{(a)} \includegraphics[angle=0,width=0.4\textwidth]{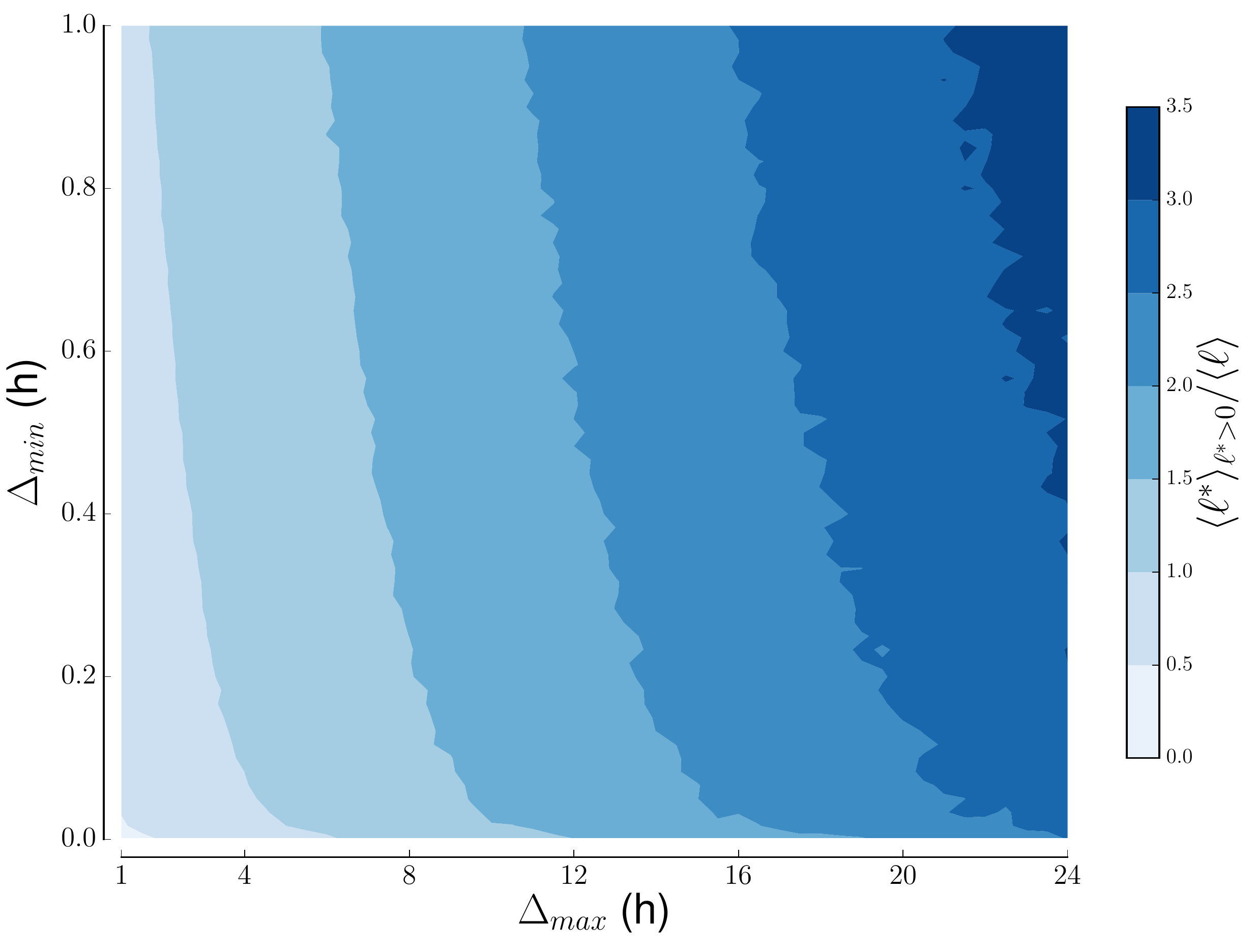}&
\raisebox{2.3cm}{(b)} \includegraphics[angle=0,width=0.4\textwidth]{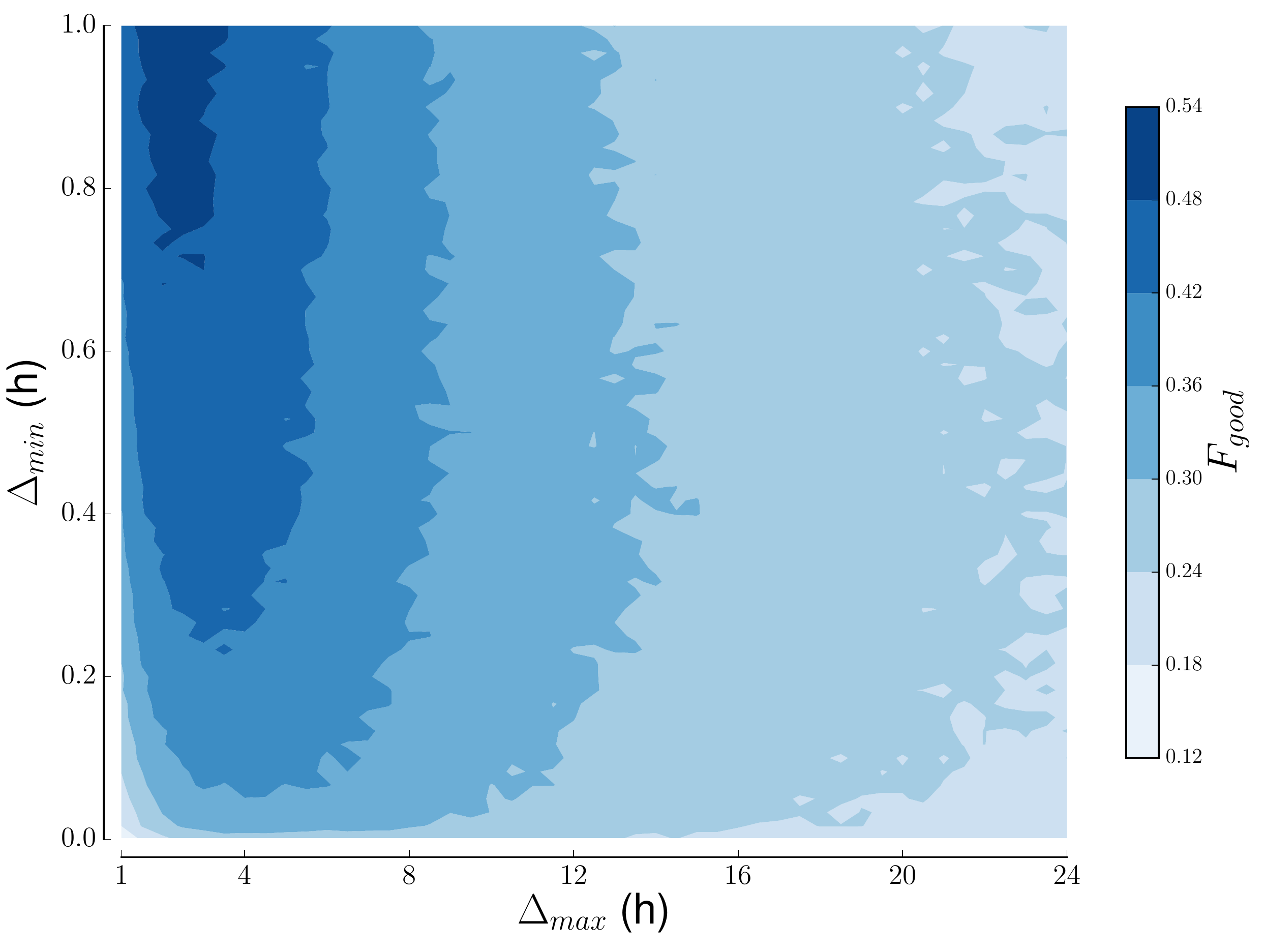}\\
\raisebox{2.3cm}{(c)} \includegraphics[angle=0,width=0.4\textwidth]{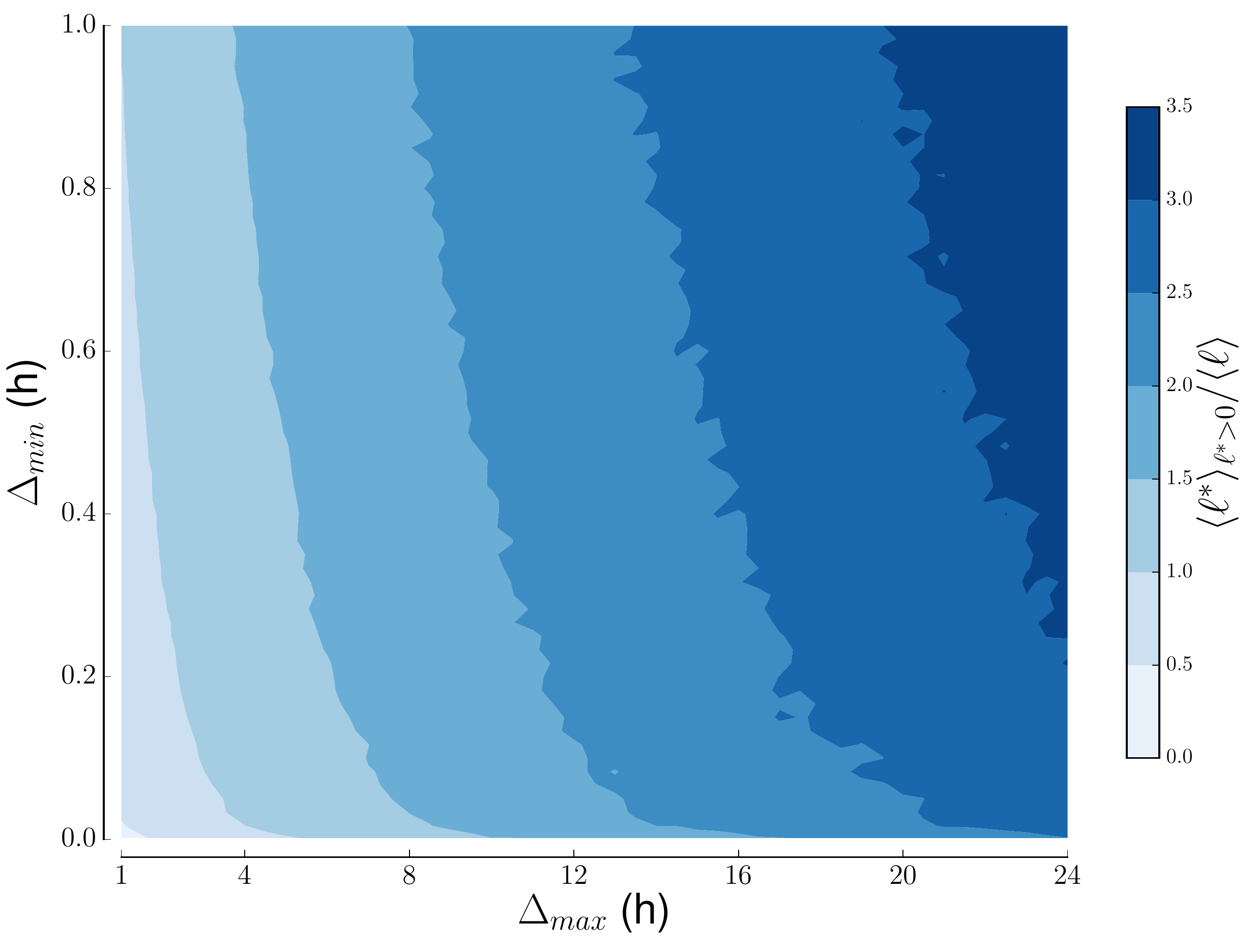}&
\raisebox{2.3cm}{(d)} \includegraphics[angle=0,width=0.4\textwidth]{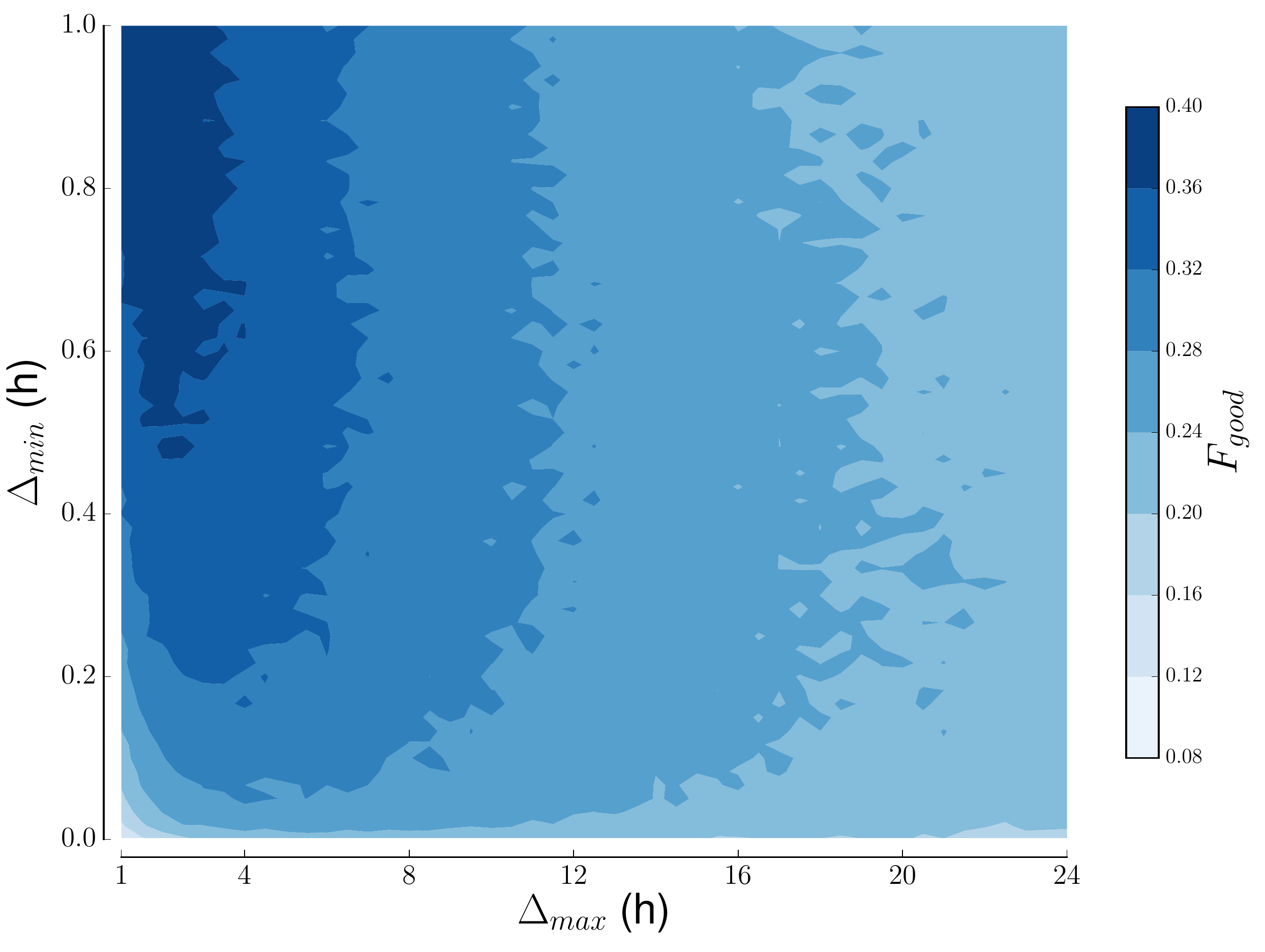}\\
\raisebox{2.3cm}{(e)} \includegraphics[angle=0,width=0.4\textwidth]{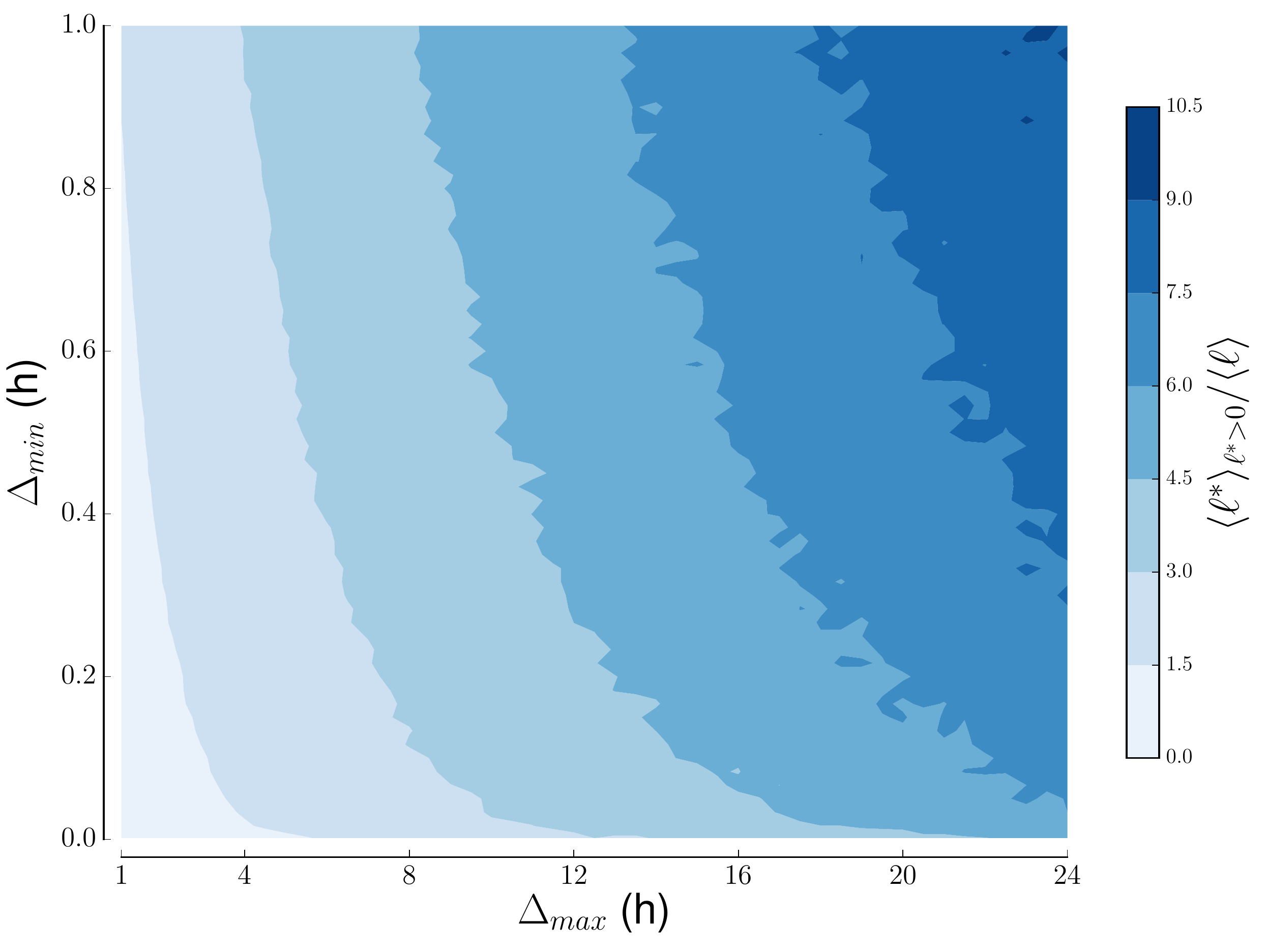}&
\raisebox{2.3cm}{(f)}  \includegraphics[angle=0,width=0.4\textwidth]{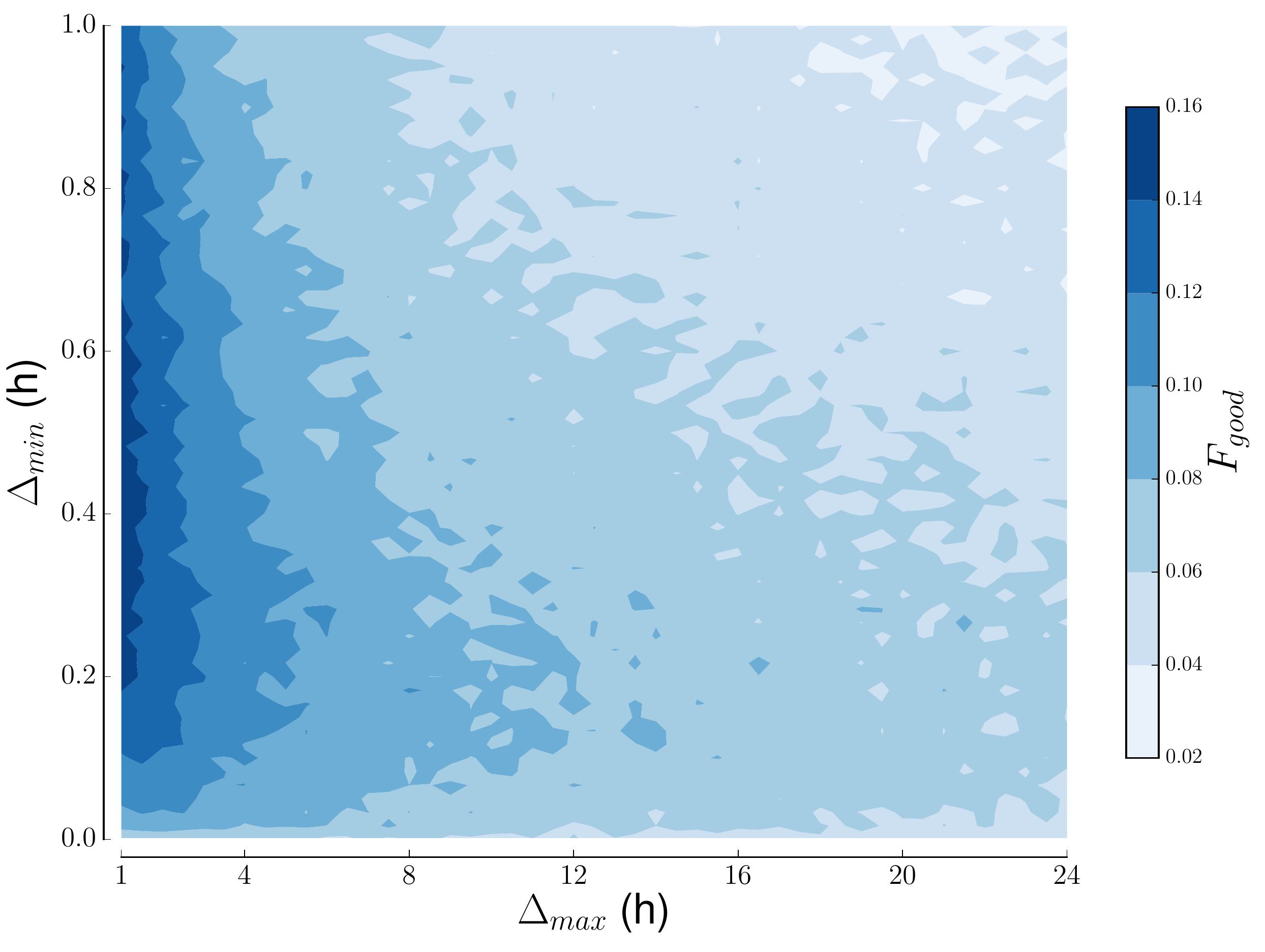}\\
\end{tabular}
\end{center}
\caption{{\bf Effect of sampling with power-law inter-event times and different rest distributions.}  Using
    sampling times distributed similarly to communication patterns, samplings
    get very bad for Exponential {\bf (a, b)}, Stretched Exponential {\bf (c, d)} or Truncated Power Law {\bf (e, f)} rest
    distributions. The quality of the sampling depends on how we choose the minimal
    $\Delta_{\rm min}$ and maximal $\Delta_{\rm max}$ inter-event times. The left panels show the error in the estimate of the average $\langle \ell\rangle$. The right panels represent the fraction $F_{\rm good}$ of correctly sampled moves.  With a very
    conservative choice of $\Delta_{\rm min} = 5$~min and $\Delta_{\rm max}=12$~h, the value $F_{\rm good}$ for the exponential rest distribution drops from
    $\approx 51\%$ to $\approx 27\%$.  For long-tailed rests we are
    limited to a maximal $F_{\rm good}\approx 23\%$ (panel (d), Stretched Exponential) when sampling human
    trajectories with mobile phones. The value drops below $6\%$ in the same conditions for the Truncated Power Law (panel (f)).}
\label{fgood_scenarios_2}
\end{figure*}

In Figs.~\ref{fgood_scenarios_1} and~\ref{fgood_scenarios_2}, we combine different rest distributions and
sampling time distributions. We see that, in all the possible scenarios,
$F_{\rm good}$ is below the best value $\hat F_{\rm good}=51\%$. In Fig.~\ref{fgood_scenarios_1}  (right) we show
the fraction $F_{\rm good}$ for exponential (exp), TPL or SE rest time distributions and with
sampling times $\Delta$ distributed as a delta function ($P(\Delta) =
\delta(\Delta-\dbar)$) or an exponential distribution ($P(\Delta) = (1/\dbar)
\exp(-\Delta/\dbar)$), associated to a Poisson process where  a sampling can happen at each moment with uniform rate. The solid blue line, together with the gray circle and
the up triangle markers, represent the same information as displayed in
Fig.~2 (d). All the other curves for $F_{\rm good}(\dbar)$ are
bell-shaped, their maxima have heights $< 51\%$,
and these maxima are reached for a $\dbar$ close to the
expected $\hat\Delta$. Therefore, all variations introduced here only yield worse
samplings. With a down triangle we show that, when the rest times are
distributed as a Stretched Exponential, as suggested by car mobility data,
the optimal sampling time would be $\check \Delta \approx 1.33$~h, but with only 39\% of
trips correctly sampled. If rest times have instead a TPL distribution,
as estimated by mobile phone data~[4], the sampling is very poor, with
$F_{\rm good}< 15\%$. 

Panels (b) (d) and (f) of Fig.~\ref{fgood_scenarios_2} correspond to a power-law sampling,
where the final result is of course dependent upon $\Delta_{\rm min}$ and
$\Delta_{\rm max}$. $F_{\rm good}$ never goes over the optimal value found for constant
sampling, which is naturally reached when $\Delta_{\rm min}$ and
$\Delta_{\rm max}$ get close to $\hat\Delta$. However, since in an individual
behavior we can easily have a whole day without communication, the values of
$F_{\rm good}$ we expect would be more of the order of the values reached on the
right half of each contour graph. We consider as reference values (here as for our analysis of the GeoLife trajectory)  $\Delta_{\rm min} =  5$~min and $\Delta_{\rm max} = 12$~h. This very conservative choice yields $F_{\rm good} = 27\%$ for the exponential rest distribution.

For both constant and random sampling times, the SE distribution yields values of $F_{\rm good}$ better than the TPL. This is expected, since larger values of $\taubar$ are associated to a better sampling (see Fig.~4 (top)), and confirms our choice of the characteristic times for vehicular mobility as the best-case scenario for our study. We therefore identify as `optimistic' values for $F_{\rm good}$ for a realistic distribution of rests (the Stretched Exponential) the value $39\%$ for constant sampling and $23\%$ for power-law sampling. This last value is, again, computed with $\Delta_{\rm min} =  5$~min and $\Delta_{\rm max} = 12$~h.

In conclusion of this section, we cannot help but remark the incongruence between the rest time distribution identified from CDR data (the TPL studied here) and the sampling time of $1$~h used to identify mobility patterns. For such a rest time distribution we predict $\langle \ell^\ast\rangle/\overline\ell \approx 2$ for a sampling time of $1$~h.  This ratio suggests that the trajectory is largely under-sampled, with only about half the trips correctly identified. Since rests would also be consequently under-counted, and thus over-estimated in duration, the rest time distribution estimated cannot be correct either.

%\newpage
\subsection{Effect of velocity and spatial embedding}

%%%%%%
\begin{figure*}[ht!]
\begin{center}
\begin{tabular}{cc}
\includegraphics[angle=0,width=0.45\textwidth]{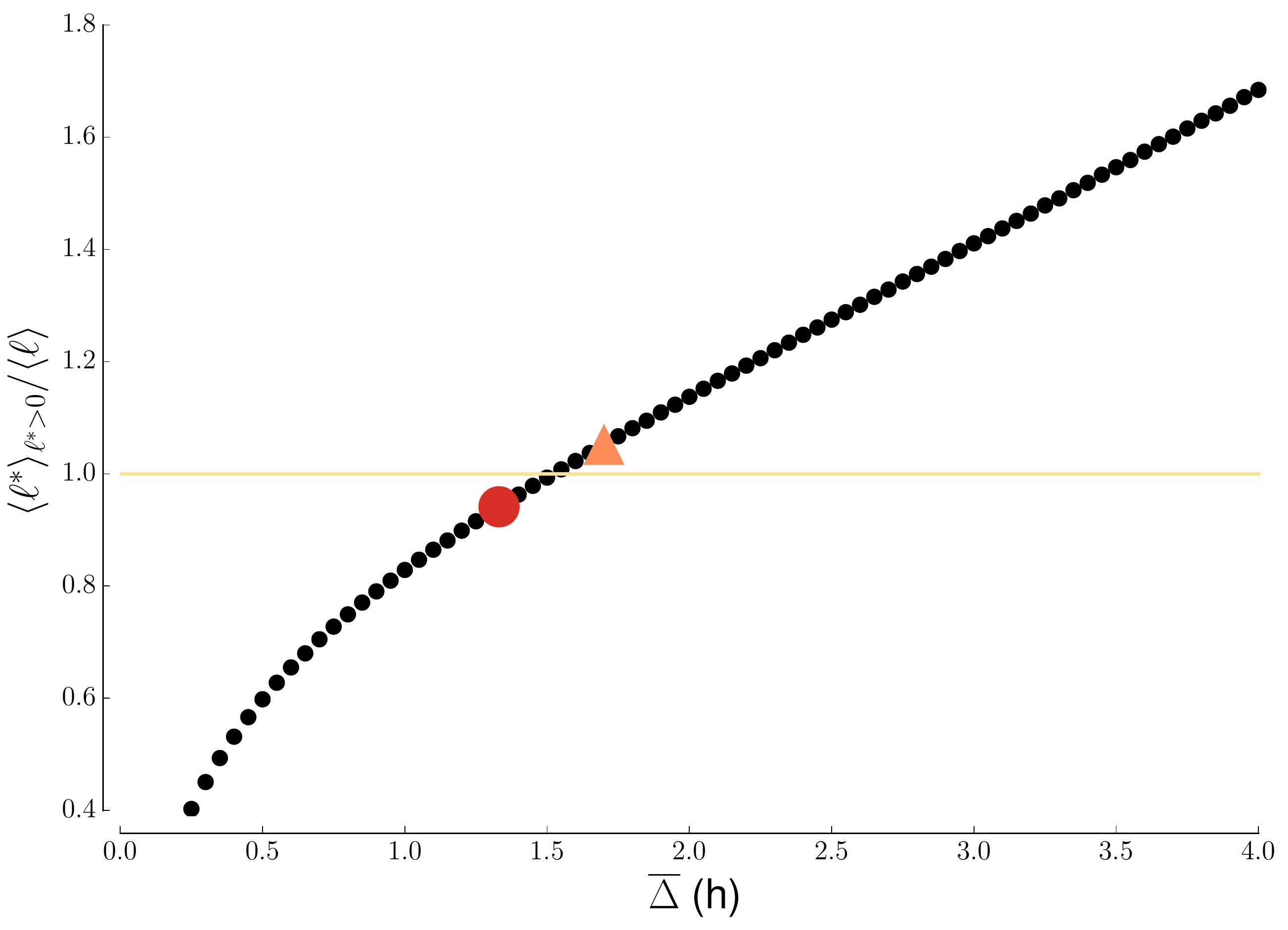}&
\includegraphics[angle=0,width=0.45\textwidth]{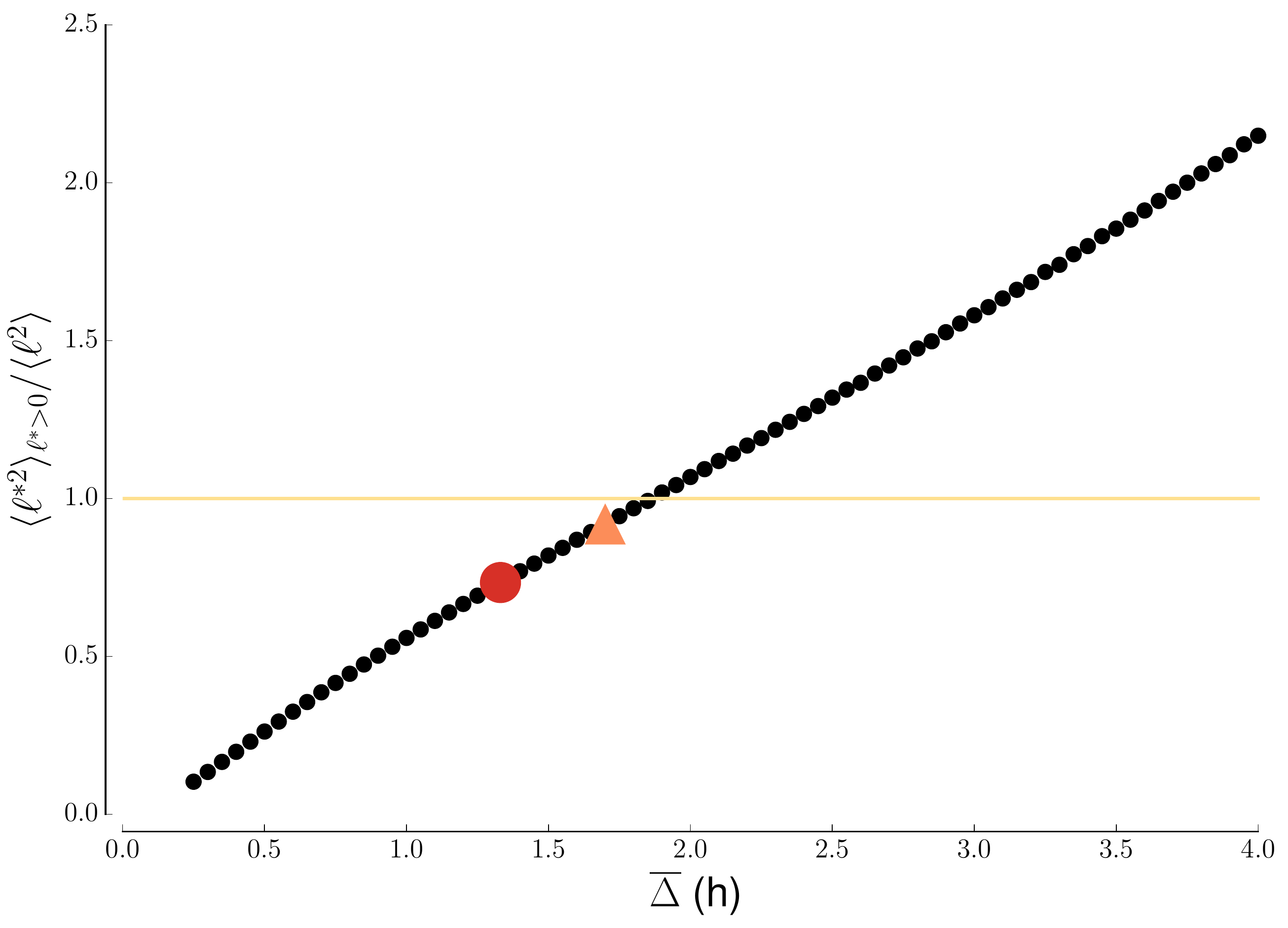}\\
\end{tabular}
\end{center}
\caption{{\bf Moments of $P(\ell^\ast)$ with speed variability in the peaked scenario.}
{\bf (Left)} First moment.
{\bf (Right)} Second moment.
We introduce variability in the speed distribution $P(v)$,
using a random acceleration model~[7]. This introduces only minor changes in estimating the moments. The red circle indicates the result using the
    sampling time $\mathring\Delta$ that matches the mean values
    ($\langle \ell^\ast\rangle = \overline\ell$). The orange triangle shows the values associated to
    $\hat\Delta$, the sampling time maximizing $F_{\rm good}$. The first sampling time $\mathring\Delta$ yields a slightly under-estimated first and second moment, with a deviation of $5\%$ and $20\%$ respectively, with respect to the moments of the displacement distribution in the
    simulated trajectories (yellow solid line). }
\label{mom_acc}
\end{figure*}
%%%%%%%%%%%%

We have seen that the optimal sampling times defined in the main text do not depend upon
the nature of $P(v)$, nor on the dimensionality of space (and thus of the vector speed $v$). Even when
introducing these two factors, the fraction of moves that have been correctly identified remains unchanged. Nevertheless, the shape of the distribution $P(\ell^\ast)$ of sampled distances necessarily depends on speed and spatial embedding.
We illustrate this by using, on top of the conditions set by Eqs.~(1), (2) and (3), a random acceleration mobility model that induces a
correlation between travel time and speed consistent with real data at a
national level~[7]. The results of Fig.~\ref{mom_acc} are to be compared with those of Fig.~3 (top) in the main text. In this case, the sampling time $\hat\Delta$ that is expected to match the first moment for constant speed under-estimates the average displacement by about $5\%$.

\begin{figure*}[ht!]
\begin{center}
 \includegraphics[angle=0,width=0.5\textwidth]{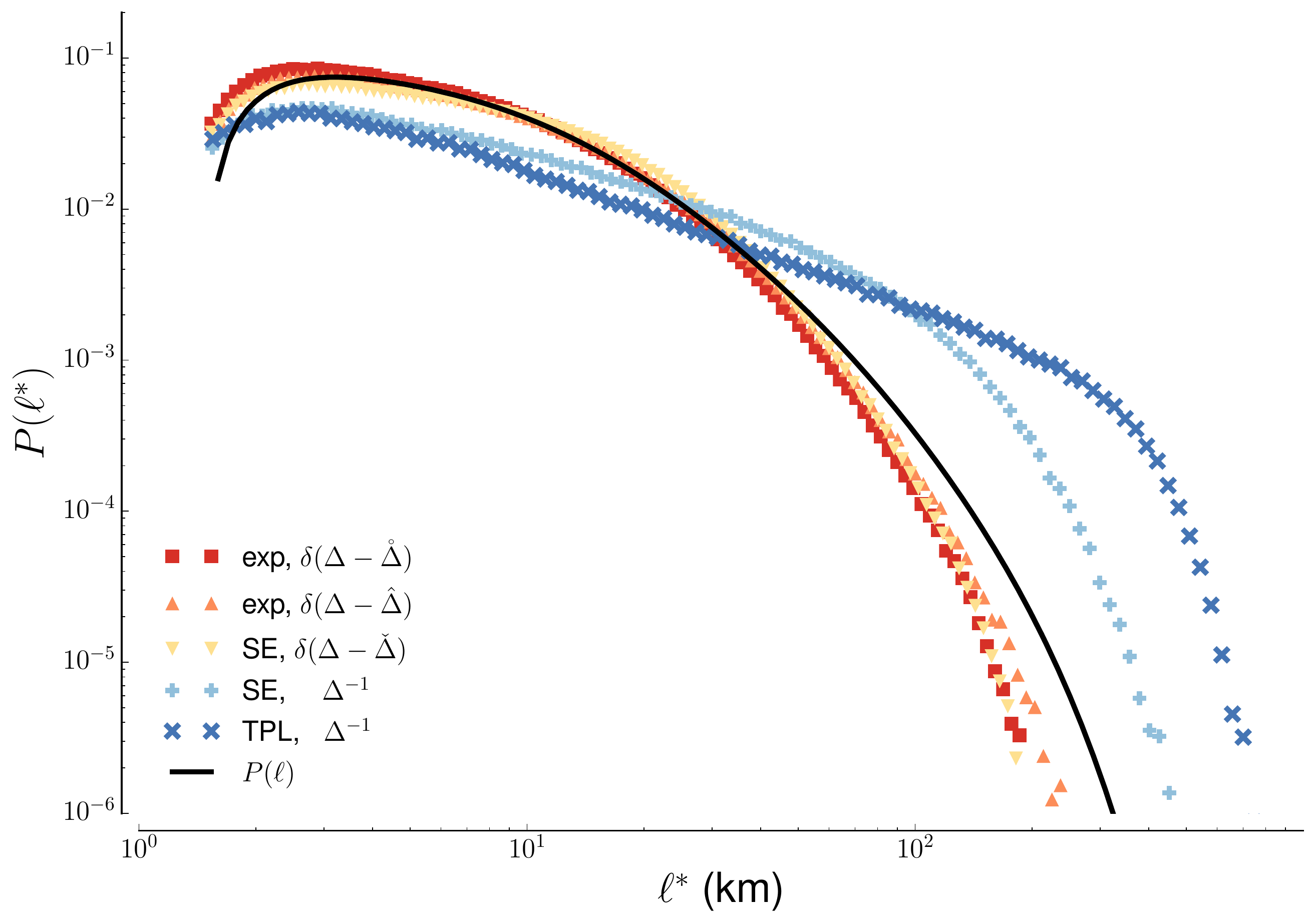}
\end{center}
\caption{
{\bf Sampling accelerated trajectories.}
    Using a random acceleration model~[7],
    we study how the displacement
    distribution changes with the sampling. The result depends upon
    the rest time distribution (exponential (exp), stretched
    exponential (SE) and truncated power law (TPL)) and on the sampling
    time distribution (peaked $\delta(\Delta-\bar\Delta)$ or with a long tail (PL)). The optimal
    sampling times ($\hat \Delta$ and $\mathring \Delta$, defined in the main text and  $\check\Delta$ defined in the preceding section) give reasonably good results, while long-tailed
    sampling creates sizeable deviations in the displacement distribution.
  }
\label{Pl_momL}
\end{figure*}

In Fig.~\ref{mom_acc} (left) we show the effect of sampling trajectories with speed
variability. We observe that, when the sampling time
distribution is broad, we have the larger deviations from the original
distribution. This same phenomenon can also be seen in Fig.~\ref{fgood_scenarios_1} (left),
where we show that the mean value $\langle \ell^\ast\rangle$ is significantly
larger when the rest distribution is broad. The issue comes by the fact that
with broad distributions we have several instances where the sampling time is
large with respect to the average rest, and therefore more jumps are joined
together.  When this happens, the dimensionality and the nature of the turning
angle distribution becomes important. Fig.~\ref{Pl_momL} would differ significantly
by introducing this further element. Indeed, at low frequencies one would have to
integrate over many re-orientations~[26]. These re-orientations are
neglected in our one-dimensional picture, that as a consequence over-estimates
this sum. A possible multi-dimensional model could be the worm-like
chain. On top of this, since the human tendency of returning
home limits the space
explored~[3], the size of these summed displacements could be
further reduced.

%%%%%%

%\newpage
\subsection{Correlations between calls and rests in empirical sampling}

In the main text we study how the statistical properties of GPS trajectories of
individuals are affected when we sample them with an inter-call distribution
extracted from mobile phone data (CDR). It can however be objected that the
times at which calls are made are correlated to the rests. Hence the
fraction of correctly sampled trips might be higher than the one we obtain
without correlations.

Moreover, more refined method of trajectory extraction are based on the idea of identifying stays (where the user is performing an activity) and pass-by's (locations where the call is made during a travel)~[40]. Normally, one needs at least two calls in the same stay to identify it correctly as an activity. The goal of this method  consists in filtering out calls made during rests. An ideal algorithm that perfectly identifies calls done during moves would then be equivalent to a perfect correlation between rests and calls.

For these reasons, we study the effect of correlations on the fraction
of correctly sampled trajectories. We introduce a parameter $p$ to quantify the
extent of correlations between calls and rests. Any call that is performed during a move is excluded with probability $1-p$.
When $p=0$ calls and rests are decorrelated, while they are perfectly correlated (all calls necessarily happen when at
rest) when $p=1$.

\begin{figure*}[ht!]
\begin{center}
 \includegraphics[angle=0,width=0.5\textwidth]{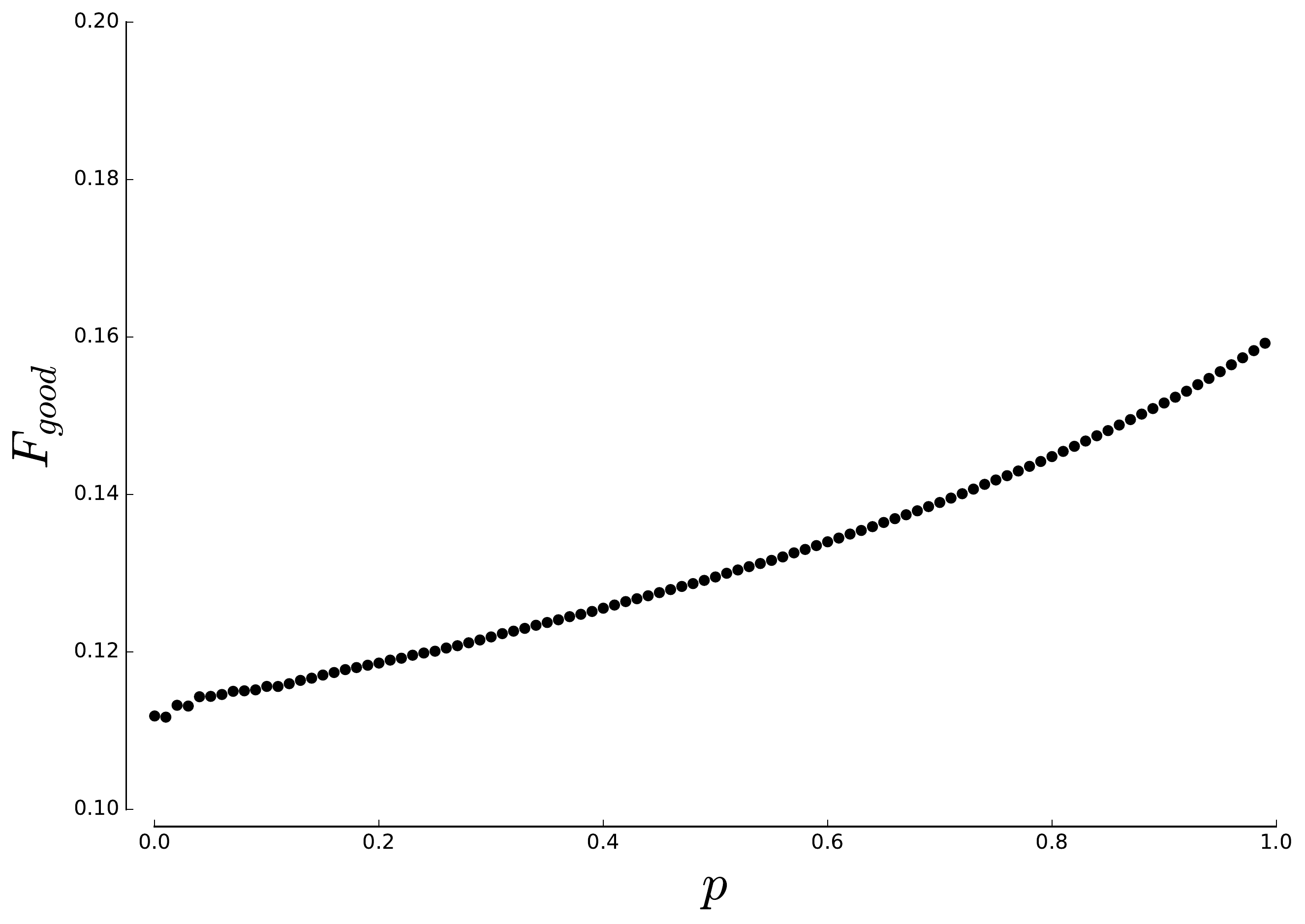}
\end{center}
\caption{
{\bf Effect of correlations on the fraction of correctly sampled trajectories.}
    We quantify the effect of correlations between calls and rests on the
    fraction of rightly sampled trips $F_{\rm good}$. When $p=0$ calls and rests are
    uncorrelated and we find $F_{\rm good} = 11\%$, as shown in the main text. When
    $p=1$ calls only happen during rests and we find $F_{\rm good} =
    16\%$. Our conclusions therefore hold disregarding of whether there are correlations between calls and movements, or if this correlation is induced by filtering out calls done during moves.
  }
\label{SI_correlations}
\end{figure*}

The results are presented on Fig.~\ref{SI_correlations}: when $p=0$ we find the
same value $F_{\rm good} = 11\%$ as presented in the main text. When $p=1$ we find
$F_{\rm good}= 16\%$, which is indeed better than without correlations, yet not
large enough to invalidate our conclusions, nor to support the current `stay point identification' method as sufficient for reconstructing mobility patterns.

% Sample bibliography item in PNAS format:
%% \bibitem{in-text reference} comma-separated author names up to 5,
%% for more than 5 authors use first author last name et al. (year published)
%% article title {\it Journal Name} volume #: start page-end page.
%% ie,
% \bibitem{Neuhaus} Neuhaus J-M, Sitcher L, Meins F, Jr, Boller T (1991)
% A short C-terminal sequence is necessary and sufficient for the
% targeting of chitinases to the plant vacuole.
% {\it Proc Natl Acad Sci USA} 88:10362-10366.

%% Enter the largest bibliography number in the facing curly brackets
%% following \begin{thebibliography}

%\newpage

\end{document}